\documentclass{aastex631}
\pdfoutput=1
\usepackage[T1]{fontenc}
\usepackage{CJK}
\usepackage{amsmath}

\shorttitle{Detectability of solar CMEs}
\shortauthors{Yang et al.}

\begin{document}

\title{Can we detect coronal mass ejections through asymmetries of Sun-as-a-star extreme-ultraviolet spectral line profiles?}

\correspondingauthor{Hui Tian, Xianyong Bai}
\email{huitian@pku.edu.cn, xybai@bao.ac.cn}

\begin{CJK*}{UTF8}{gbsn}
\author[0000-0002-4973-0018]{Zihao Yang（杨子浩）}

\affiliation{School of Earth and Space Sciences, Peking University, Beijing 100871, China}

\author[0000-0002-1369-1758]{Hui Tian（田晖）}
\affiliation{School of Earth and Space Sciences, Peking University, Beijing 100871, China}

\author[0000-0003-2686-9153]{Xianyong Bai}
\affiliation{National Astronomical Observatories, Chinese Academy of Sciences, Beijing 100012, China}

\author[0000-0001-5494-4339]{Yajie Chen}
\affiliation{School of Earth and Space Sciences, Peking University, Beijing 100871, China}

\author[0000-0002-9293-8439]{Yang Guo}
\affiliation{School of Astronomy and Space Science, Nanjing University, Nanjing 210023, China}
\affiliation{Key Laboratory of Modern Astronomy and Astrophysics (Nanjing University), Ministry of Education, Nanjing 210023, China}

\author[0000-0003-3908-1330]{Yingjie Zhu（朱英杰）}
\affiliation{Department of Climate and Space Sciences and Engineering, University of Michigan, Ann Arbor, MI 48109, USA}

\author[0000-0003-2837-7136]{Xin Cheng}
\affiliation{School of Astronomy and Space Science, Nanjing University, Nanjing 210023, China}
\affiliation{Key Laboratory of Modern Astronomy and Astrophysics (Nanjing University), Ministry of Education, Nanjing 210023, China}

\author[0000-0002-6641-8034]{Yuhang Gao（高宇航）}
\affiliation{School of Earth and Space Sciences, Peking University, Beijing 100871, China}
\affiliation{Centre for mathematical Plasma Astrophysics (CmPA), KU Leuven, Celestijnenlaan 200B bus 2400, B-3001 Leuven, Belgium}

\author[0000-0002-7421-4701]{Yu Xu（徐昱）}
\affiliation{School of Earth and Space Sciences, Peking University, Beijing 100871, China}

\author[0000-0001-7866-4358]{Hechao Chen}
\affiliation{School of Earth and Space Sciences, Peking University, Beijing 100871, China}

\author{Jiale Zhang}
\affiliation{School of Earth and Space Sciences, Peking University, Beijing 100871, China}



\begin{abstract}

Coronal mass ejections (CMEs) are the largest-scale eruptive phenomena in the solar system. Associated with enormous plasma ejections and energy release, CMEs have an important impact on the solar-terrestrial environment. Accurate predictions of the arrival times of CMEs at the Earth depend on the precise measurements on their three-dimensional velocities, which can be achieved using simultaneous line-of-sight (LOS) and plane-of-sky (POS) observations. Besides the POS information from routine coronagraph and extreme ultraviolet (EUV) imaging observations, spectroscopic observations could unveil the physical properties of CMEs including their LOS velocities. We propose that spectral line asymmetries measured by Sun-as-a-star spectrographs can be used for routine detections of CMEs and estimations of their LOS velocities during their early propagation phases. Such observations can also provide important clues for the detection of CMEs on other solar-like stars. However, few studies have concentrated on whether we can detect CME signals and accurately diagnose CME properties through Sun-as-a-star spectral observations. In this work, we constructed a geometric CME model and derived the analytical expressions for full-disk integrated EUV line profiles during CMEs. For different CME properties and instrumental configurations, full disk-integrated line profiles were synthesized. We further evaluated the detectability and diagnostic potential of CMEs from the synthetic line profiles. Our investigations provide important constraints on the future design of Sun-as-a-star spectrographs for CME detections through EUV line asymmetries.

\end{abstract}



\section{Introduction} \label{sec:intro}

Coronal mass ejections (CMEs) are the largest-scale eruptive events in the solar system. During these events, a large amount of magnetized plasma is expelled from the Sun into the interplanetary space, and may interact with the solar system planets. CMEs play a dominant role in driving disturbances of the solar-terrestrial space environment, and are among the primary sources of severe space weather phenomena in the solar system \citep[e.g.,][]{2016GSL.....3....8G}. 

Remote sensing observations of CMEs are primarily achieved through white-light coronagraphs or EUV imagers. Routine observations of solar CMEs are usually obtained  using white-light coronagraphs such as the Large Angle Spectrometric Coronagraph \citep[LASCO,][]{1995SoPh..162..357B} onboard Solar and Heliospheric Observatory (SOHO) and K-coronagraph \citep[K-Cor,][]{2012SPIE.8444E..3ND} of Coronal Solar Magnetism Observatory (COSMO), and EUV imagers such as the Atmospheric Imaging Assembly \citep[AIA,][]{2012SoPh..275...17L} oboard Solar Dynamics Observatory (SDO) \citep[e.g.,][]{2003A&A...397.1057Z,2003ApJ...588L..53G,2009EM&P..104..295G,2017SpWea..15..240S,2012ApJ...761...62C,2020ApJ...894...85C}. These observations can provide rich information on the kinematics of CMEs. However, what they observe are just the plane-of-sky (POS) projections of CMEs. Accurate predictions of CME kinematics and its impact on the solar-terrestrial space environment require information on its real speed and propagation direction.  Spectral observations can provide the line-of-sight (LOS) velocity through Doppler effect. A combination of POS propagations from traditional white-light coronagraphs/EUV imagers with LOS propagations revealed through spectral observations can help the reconstruction of 3D kinematics of CMEs at early propagation stages. Spectral observations may be achieved through tunable-filter coronagraphs such as Coronal Multi-channel Polarimeter \citep[CoMP,][]{2008SoPh..247..411T}, which could detect the LOS velocity through imaging spectroscopy \citep[e.g.,][]{2013SoPh..288..637T}, but are limited to off-limb observations. EUV slit spectrographs can also reveal the LOS propagations of CMEs.  However, due to the limited field-of-view (FOV) of a slit spectrograph, the possibility of capturing CMEs is extremely low. As a result, spatially resolved spectroscopic observations of solar CMEs are still rare. Only very few studies  using EUV Imaging Spectrometer \citep[EIS,][]{2007SoPh..243...19C} onboard the Hinode satellite have successfully observed CME-related eruptions \citep[e.g.,][]{2010ApJ...711...75L,2012ApJ...748..106T}. In addition, Ultraviolet Coronagraph Spectrometer (UVCS) onboard SOHO has captures some CME eruptions in off-limb regions, but mainly using cooler transition region lines formed at a temperature of about $10^{5.0}\ $K \citep{2013JGRA..118..967G}. Moreover, its diagnosing capability was also limited by the 
field-of-view of the slit. With these difficulties, routine spectral observations of solar CMEs are not available yet.

Alternately, Sun-as-a-star spectroscopic observations have the potential to reveal global plasma properties on the Sun without limitations of FOV. In this case, LOS signals contributed by Earth-facing CMEs could be observed, increasing the probability of capturing CMEs. 
With required instrumental sensitivities, full disk-integrated spectral observations could provide an alternative way to routine monitoring of CMEs. From spatially resolved spectra such as those obtained from Hinode/EIS, a blue-shifted secondary spectral component caused by ejections can be observed, resulting in blueward asymmetries of spectral line profiles \citep[e.g.,][]{2009ApJ...701L...1D,2011ApJ...732...84M,2011ApJ...738...18T}. It is expected that such a secondary component and line asymmetries might also be observed from full-disk integrated spectra. Using spectral line asymmetries from Sun-as-a-star observations, the current routine observations on POS components of CME velocities will be complemented with their derived LOS counterparts. Recently, \cite{Xu2022} combined the Sun-as-a-star spectroscopic observation of cooler transition region lines from Extreme-ultraviolet Variability Experiment \citep[EVE,][]{2012SoPh..275..115W} onboard SDO and the imaging observation from Solar Terrestrial Relation Observatory \citep[STEREO,][]{2008SSRv..136....5K}, and derived the full velocity of the bulk motion of a mass ejection. However, the possibility of CME detections using Sun-as-a-star spectral line asymmetries under different solar conditions and instrumental configurations still needs to be assessed. 

Whereas solar CMEs can be routinely observed using coronagraphs/EUV imagers, it is still very difficult to detect signals of their stellar counterparts. As mentioned before, LOS propagation of solar CMEs can result in line asymmetries in spatially-resolved spectra \citep{2012ApJ...748..106T}. Similar to Sun-as-a-star observation, such asymmetries could also be possibly detected in stellar spectra. Using X-ray spectral observations of a giant star, \cite{2019NatAs...3..742A} discovered the blue-shifted components in the line profiles, which they explained as being caused by a stellar CME associated with a strong flare. The LOS propagation of a large amount of plasma from stellar CMEs will contribute to a secondary blue or red-shifted component of the spectral line profiles, leading to spectral line asymmetries. Nevertheless, among the very few identified stellar CME candidates, most of them are discovered through line asymmetries using lines formed at chromospheric or transition region temperatures such as H$\alpha$, H$\beta$, H$\gamma$, O \sc{iv}\rm{} or C \sc{iii}\rm{} \citep[e.g.,][]{1990A&A...238..249H,2011A&A...536A..62L,2019A&A...623A..49V,2021NatAs...6..241N}. These observations could reveal the information of the cooler plasma within CMEs. In contrast, the spectral lines with formation temperatures of $10^{5.8}-10^{6.3}\ $K usually observed in EUV wavelengths are better tracers for coronal plasma in CMEs.
Therefore, spectral observations utilizing these EUV lines may be promising in the discovery of stellar CMEs in the solar neighborhood. Nonetheless, this remains barely explored yet. When the Sun is observed as a star, we can evaluate the possibility of solar CME detections using the asymmetries of full-disk integrated spectral profiles. Similarly, such studies may also serve as important clues for the detection of stellar CMEs on other solar analogs.

In this work, starting with the spatially resolved spectral observations of a solar CME region obtained in EUV wavelengths, we developed a CME model, and derived the analytical formalism of Sun-as-a-star line profiles during CMEs. Based on the analytical expression, we synthesized the EUV spectral line profiles of solar CMEs at different evolutionary stages. We varied the solar activity levels (the areas of CME regions and active regions) and instrumental parameters such as signal-to-noise ratio (SNR) and spectral resolution, and generated different full disk-integrated spectral line profiles. Furthermore, we used a quantitative criterion to investigate the detectability of CMEs from synthetic full disk-integrated line profiles. In order to assess the accuracy of LOS velocities and peak intensities of CMEs determined from the synthetic profiles, we compared these values with the input ones for different sets of instrumental parameters. Our investigations provide important constraints for the design of future potential Sun-as-a-star spectrographs. Our results may also provide unique references for similar detections of stellar CMEs on solar-like stars.

\section{Methodology} \label{sec:methods}
In this section, we will focus on the methodology for the synthesis of full disk-integrated spectral line profiles. We describe our observations in Sect. \ref{obs}. In Sect. \ref{models}, we introduce our kinematics and geometric models of CMEs, as well as the assumptions made during the calculations. The derivations of spectral line emissions and profiles are described in Sect. \ref{emission}. Finally, we investigate the effects of CME expansions on spectral line profiles in Sect. \ref{expansion}.
\subsection{Observations}\label{obs}

\begin{figure}[htbp!]
\centering
\includegraphics[width=0.9\textwidth]{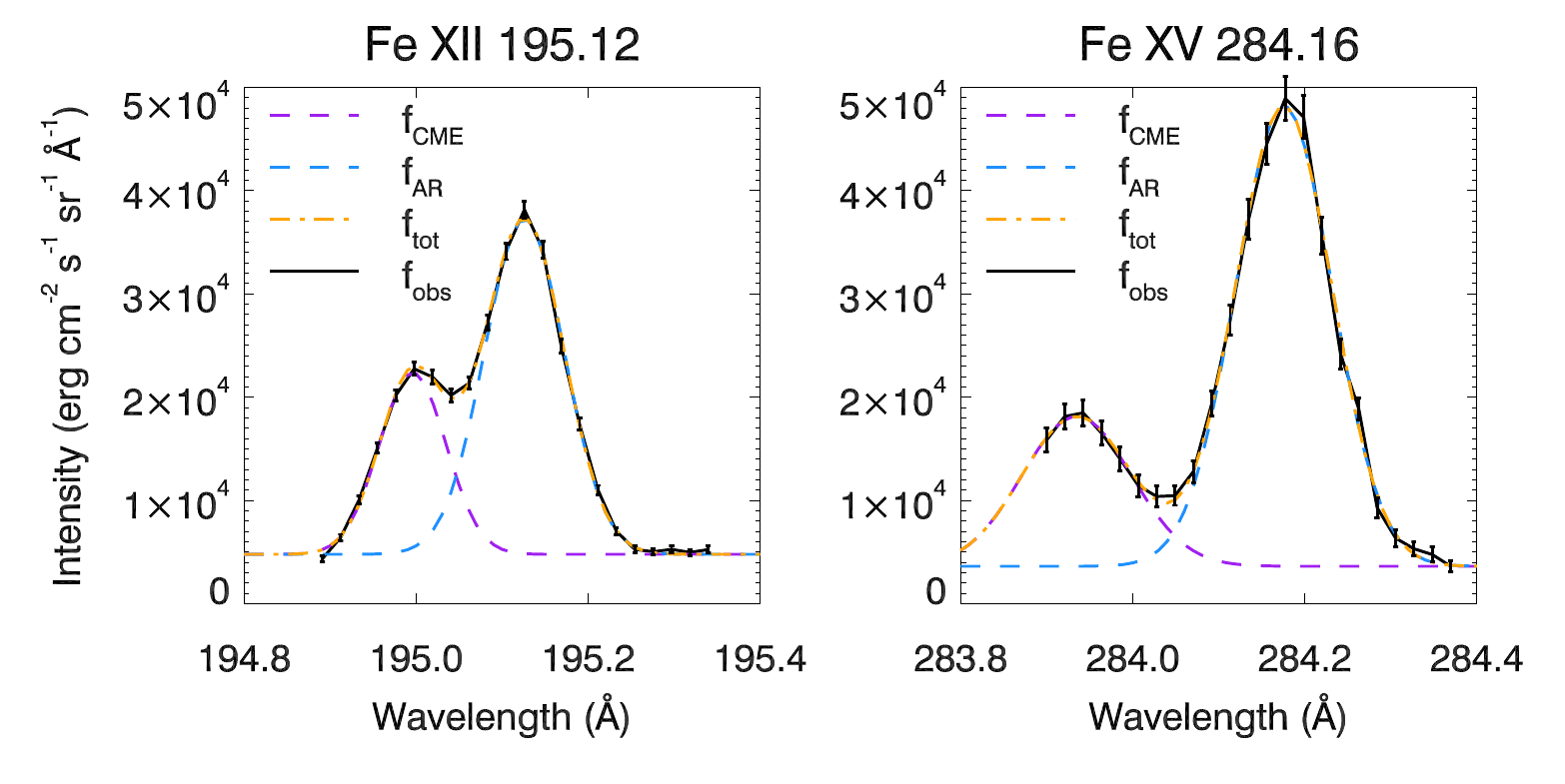}
\caption{The line profiles of Fe \sc{xii}\rm{} 195.12 \AA\ (left) and Fe \sc{xv}\rm{} 284.16 \AA\ (right) averaged over the erupted loop region observed by Hinode/EIS. The black solid profiles with error bars are the observed profiles averaged over the small erupted region. The purple dashed profiles are the Gaussian fits to the blue-wing components caused by erupted plasma, the blue dashed profiles are the Gaussian fits to the primary components emitted from background active region, and the orange dot-dashed profiles are the double Gaussian fits to the observed profiles.}  \label{fig:f1_obs_CME_profile}
\end{figure}

Before synthesizing the Sun-as-a-star line profiles, we used the EUV spectral observation of an erupted loop from Hinode/EIS as our reference. The observation was made using a $2\arcsec\times 160\arcsec$ slit scan from 19:13 UT to 20:06 UT on Feb. 14, 2011. This observation was first analyzed by \cite{2012ApJ...748..106T}. During this observation, several EUV spectral lines were observed. The Fe \sc{xii}\rm{} 195.12 \AA\ and Fe \sc{xv}\rm{} 284.16 \AA\ lines are among the strongest lines, and their formation temperatures are close to and higher than the typical coronal temperature, respectively. These two lines are also the cleaner lines without severe line blending. Therefore, we choose these two lines for analysis. We applied double-Gaussian fits to the Fe \sc{xii}\rm{} 195.12 \AA\ and Fe \sc{xv}\rm{} 284.16 \AA\ line profiles averaged over a small mass ejection region of $2\arcsec\times 5\arcsec$. Figure \ref{fig:f1_obs_CME_profile} shows the observed line profiles and the associated Gaussian fittings. Each fitted profiles has two Gaussian components: a strong primary peak, corresponding to the emission from the background active region, and a blue-wing component caused by the erupting plasma. Hereafter, we will use the primary Gaussian components as the reference Fe \sc{xii}\rm{} 195.12 \AA\ and Fe \sc{xv}\rm{} 284.16 \AA\ profiles for typical solar active regions without eruptions (with the intensity of $\bar{I}_{\text{AR}}$), and the blue-wing profiles will be used as the reference profiles for the ejected plasma of CME at the observed height $h_0$ (the intensity is $\bar{I}_{\text{CME}}(h_0)$). From EIS observations in active regions (ARs) and quiet Sun (QS) regions, we find that the line intensity ratios in AR and QS will fluctuate over a range of values. In this case, we assumed that the Fe \sc{xii}\rm{} 195.12 \AA\ line intensity in active regions is about 9 times stronger than that in quiet Sun regions ($\bar{I}_{\text{QS}}=\bar{I}_{\text{AR}}/9$), which falls in a reasonable range of intensity ratios according to EIS observations. For the Fe \sc{xv}\rm{} 284.16 \AA\ line, similarly, we assumed $\bar{I}_{\text{QS}}=\bar{I}_{\text{AR}}/375$. The Gaussian line profiles are expressed as a function of the wavelength $\lambda$ by $f(\lambda)=I_\text{p} \exp{\left(-(\lambda-\lambda_0)^2/\Delta\lambda_\text{D}^2\right)}+I_\text{B}$, where $I_\text{p}$ is the peak intensity, $\Delta\lambda_\text{D}$ is the 1/e line width \citep[see][]{2010A&A...521A..51P}, $I_\text{B}$ is the background continuum intensity, and $\lambda_0$ is the centroid wavelength of the spectral line. The line intensity is then written as  $\bar{I}=\sqrt{\pi}I_\text{p}\Delta\lambda_\text{D}$. Based on the selected intensity ratios in AR and QS, and assuming that the line width in quiet Sun regions is the same as the one in active regions without CME eruptions, we then generated the reference line profiles in quiet Sun regions.

\subsection{Models and Assumptions}\label{models}
In order to derive the analytical expression for Sun-as-a-star line profiles during CMEs, the information on the kinematics and plasma properties at the early stages of CMEs are required. Additionally, due to the expansion of CME structures as it propagates, such effect on synthetic line profiles also need to be included. In the following we will derive the analytical expression based on commonly used models and reasonable assumptions.

\subsubsection{Kinematic Evolution Model at Early Stage of CME propagation}
\begin{figure}[htbp!]
\centering
\includegraphics[width=0.5\textwidth]{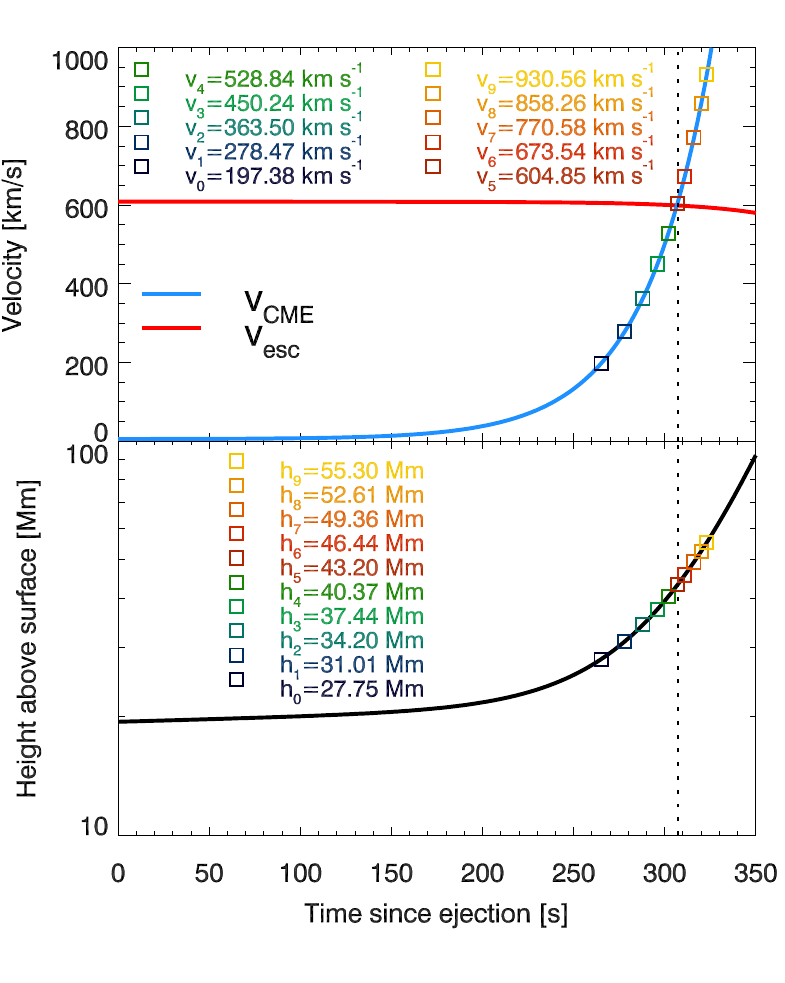}
\caption{The evolution of CME height and velocity as a function of time. In the upper panel, the blue line represents $v_{\text{CME}}$ (CME velocity from the best-fitted kinematic model), and the red line is $v_{\text{esc}}$ (the local escape velocity). In the lower panel, the black line is the CME height from the kinematic model. The vertical dashed line marks the time step where the CME is accelerated to the local escape velocity ($v_{\text{CME}}=v_{\text{esc}}$). The selected time steps used for further analysis are marked by colored squares overlaid on the velocity and height profiles.}  \label{fig:f2_cme_esc_vel_h_new}
\end{figure}

In this study, we focused mainly on the early propagation of CMEs, since the EUV emissions of CME at this stage are still strong for detections, as revealed in EUV imaging observations. Previous studies have shown that CMEs usually experience a slow upward propagation followed by a main-acceleration phase, then propagate with a nearly constant velocity \citep[e.g.,][]{2001ApJ...559..452Z,2003ApJ...588L..53G,2004ApJ...604..420Z}. \cite{2020ApJ...894...85C}
 investigated several CME eruptions using EUV imaging observations, and found the best-fitted functions to describe the early evolution of CME kinematics. Our observed CME height \citep[$\sim 30$ Mm, see Figure 5 and discussions of Section 3.3 in][]{2012ApJ...748..106T} and velocity ($\sim 200\ \text{km}\ \text{s}^{-1}$, which can be derived from Figure \ref{fig:f1_obs_CME_profile}) are consistent with values of sample H6 in \cite{2020ApJ...894...85C}. Therefore, the temporal evolution profiles of height and velocity for sample H6 were adopted as the kinematic evolution models of the CME in our analysis. Figure \ref{fig:f2_cme_esc_vel_h_new} depicts the velocity and height evolution profiles showing the characteristic of a fast-accelerated CME. To investigate the properties of full disk-integrated spectral line profiles as the CME propagates, we selected 10 different heights (from $h_0\sim 30$ Mm, which is the height where the CME was observed by EIS, to the height where the velocity reaches around $1000\ \text{km}\ \text{s}^{-1}$, with equally spaced interval) from the kinematic model for further analysis. The heights and velocities at the 10 points are referred to as $h_\text{i}$ and $v_\text{i}$ ($\text{i}=0,\ 1,\ ...,\ 9$), denoted by colored squares in Figure \ref{fig:f2_cme_esc_vel_h_new}. 
 
When studying the space weather effect of CMEs, one essential criterion is whether a CME can successfully erupt into the interplanetary space. This can be determined from the comparison between the final velocity of CMEs and the local escape velocity. A successfully erupted CME usually refers to the one eventually exceeding local escape velocity \citep[e.g.,][]{2019ApJ...877..149A}. From Figure \ref{fig:f2_cme_esc_vel_h_new}, it is found that as the CME undergoes the main-acceleration phase, its velocity will finally exceed the local escape velocity at the height and velocity of $h_5\approx 43.20\ $Mm and $v_5\approx 604\ \text{km}\ \text{s}^{-1}$, respectively. Consequently, we regarded this CME as a successful eruption. In addition to the deceleration of gravity, magnetic pressure and pressure gradient play important roles in accelerating CMEs \citep{2012JGRA..11711101S}. The combination of acceleration forces and deceleration forces will lead to a smaller critical velocity that the CME can escape into interplanetary space. Therefore, the critical velocity we derived here can be regarded as an upper limit.

\subsubsection{Geometry and Self-Similar Expansion of CME Structures}

The geometry and expansion of CME structures play important roles when synthesizing line profiles during CMEs. Conventionally, CMEs can be geometrically described by the graduated cylindrical shell (GCS) model \citep{2006ApJ...652..763T}, which treats the CME as a flux tube. However, the GCS model is more suitable for numerical simulations than analytical calculations owing to its complexity. In this work, we aim to provide an analytical expression of the line profiles based on simpler but reasonable models. Several previous works have used simplified flux tube models instead of the GCS model. In this work, based on the geometric models presented in \cite{2009JGRA..11410104W} and \cite{2020AnGeo..38..657C}, we used the Croissant-like shape model to characterize the structure of a CME. In this case, we neglected the legs in the GCS model, leaving the partial donut
structure as the primary part of the CME flux tube. In the mean time, self-similar evolution is usually a proper approximation for the expansion behavior of CMEs \citep[e.g.,][]{1984ApJ...281..392L,2006ApJ...652.1747C,2009SoPh..260..401M,2020SoPh..295..107B}. Therefore, in our model, we also assumed the CME experiencing a self-similar expansion during propagation, as depicted in the orange colored shape in Figure \ref{CME_new}(A). We will give more detailed descriptions on the effect of self-similar expansion to the observed LOS velocity of CMEs in Sect. \ref{expansion}. 

For the convenience of analytical derivation, we assumed that the CME structure has a uniform density distribution and is isothermal during its early propagation. Moreover, the CME structure is regarded as mass conserved at this stage. Although these assumptions are probably over-simplifications, they have been widely used in previous CME studies as reasonable assumptions \citep[e.g.,][]{2003ApJ...588..586D,2006ApJ...642.1216V,2009AnGeo..27.3275A,2017NatSR...7.4152O,2021JSWSC..11....8N}.

Supposing that at two different heights $h_0$ (the height where the CME was observed by EIS) and $h$ (the distance between the center of flux tube and solar surface), the radii of the CME flux tube are $r_0$ and $r$, respectively. If viewed from the north pole of the Sun (the polar view), as depicted in Figure \ref{CME_new}(A), we can define a polar field angle as $2\theta_0$. When viewed equatorially (the lateral view, see Figure \ref{CME_new}(B)), we defined a lateral field angle as $2\gamma$. We used $R_\text{s}$ to denote the solar radius.

From the lateral view in Figure \ref{CME_new}(B), we can write 
\begin{equation}
    \sin\gamma=\frac{r_0}{R_\text{s}+h_0}=\frac{r}{R_\text{s}+h}=\kappa 
\end{equation}
here we used $\kappa$ to denote $\sin\gamma$.

At the initial height $h_0$, we assumed that the radius of the flux tube is equal to the height from its center to solar surface: $r_0=h_0$. In this case, the volume of the partial donut structure at a given height $h$ is

\begin{equation}
V(h)=\pi r^2(R_\text{s}+h)\cdot 2\theta_0
\end{equation}
where $\theta_0$ is in the unit of radian.

As we mentioned earlier, the CME is assumed to be uniform in density. At the height $h$, the density is denoted by $n_\text{e}(h)$. Under the mass conservation assumption, we have 
\begin{equation}\label{eq:density}
    \dfrac{n_\text{e}(h)}{n_\text{e}(h_0)}=\dfrac{V(h_0)}{V(h)}=\dfrac{\pi r_0^2(R_\text{s}+h_0)}{\pi r^2(R_\text{s}+h)}=\dfrac{\kappa^2(R_\text{s}+h_0)^3}{\kappa^2(R_\text{s}+h)^3}=(\dfrac{R_\text{s}+h}{R_\text{s}+h_0})^{-3}
\end{equation}
This equation describes the height variation of electron density in the CME structure. A comparison with observations of electron densities in CMEs has shown very similar behavior to our derivations \citep[e.g.,][]{2016JGRA..121.2853L}.

\subsection{Spectral Line Emission Calculation}\label{emission}

In this part, we mainly dealed with the derivations of CME emissions as functions of height. In Sect. \ref{fluxdef}, we will first review the concepts of intensity (radiance) and flux (irradiance) which are fundamental for our spectral line synthesis. We then derived the height variation of spectral line emission and continuum emission of CMEs in Sect. \ref{intensityvariation} and \ref{continuum}, respectively.

\subsubsection{Intensity and Flux}\label{fluxdef}
In terms of spectral observations using Hinode/EIS, the emission we obtained is the line intensity. While for Sun-as-a-star observations, what we should synthesize is the flux of a spectral line. Before calculating the EUV emissions of a CME at different heights, we will briefly go through the basic definitions of intensity (radiance) and flux (irradiance), as well as the relationship between them.

Since most solar observations are spatially resolved, the received emission is intensity or radiance. Intensity emitted by the Sun is the energy passing through unit area (e.g., the emitting area) per unit solid angle (the solid angle is subtended by the detector pixel area with the apex on the emitting source) per unit time. The unit of intensity/radiance is $\text{erg}\ \text{cm}^{-2}\ \text{s}^{-1}\ \text{sr}^{-1}$ \citep{2012uxss.book.....P}. The intensity detected at each pixel of Hinode/EIS is numerically equal to the intensity emitted from the corresponding solar source region defined by the pixel spatial scale. In our work, we averaged the intensity in a small region where the ejection occurs. This means that the ``intensity'' we obtained can be treated as the average intensity per unit area, denoted by $\bar{I}= (\int{I\text{d}S})/S$. Here $I$ is the intensity per unit area, $\int{I\text{d}S}$ is the integration of intensity over the area of the source region, and $S$ is the surface area of the source region. This is similar to the ``average disk intensity'' introduced in \cite{1999ApJ...518..480F}. A detailed discussion is presented in the following paragraphs.

However, for Sun-as-a-star spectral observations, we do not have spatially resolved but only full disk-integrated spectra. In this case, the emission received by the detector is flux or irradiance. Irradiance is the energy passing through unit area (the detector) per unit time. It is the integral of radiance over the solid angle subtended by the surface of the source \citep[e.g.,][]{1999ApJ...518..480F,2012uxss.book.....P}. The unit of flux/irradiance is $\text{erg}\ \text{cm}^{-2}\ \text{s}^{-1}$. The magnitude of flux/irradiance is dependent on the source-receiver distance.

As discussed earlier, the observed emission from the spatially resolved Hinode/EIS is the average intensity per unit area, $\bar{I}$. While the emission we aim to synthesize for Sun-as-a-star case is the flux $F$. 
The solid angle subtended by the source is set to be normal to the detector. If the source-receiver distance is $d$ and the source area is $S$, then the solid angle subtended by the source will be $\Omega=S/d^2$. Similar to \cite{1999ApJ...518..480F}, we can write the flux as
\begin{equation}\label{eq:Fontenla_E1}
    F=\dfrac{\bar{I}\cdot S}{d^2}=\dfrac{I_\text{tot}}{d^2}
\end{equation}

For instruments in the near-Earth orbit, the source-receiver distance $d$ is nearly constant during observations. It is to be noted that throughout the work we only cared about the relative intensity. Thus, for simplification purpose, we can set $d$ to unity. In addition, we only focused on the percentage of the source area relative to the full disk area, so instead of the absolute value of $S$, we only needed to know the relative percentage. 

\subsubsection{Unit Volume Emissivity, Line Intensity and Flux}\label{intensityvariation}

In this section we will introduce the definitions of unit volume emissivity, which are related to the line intensity and flux we aim to synthesize. 

Unit volume emissivity (emissivity per unit volume) from the transition between levels $i$ and $j$ can be written as \citep{2018LRSP...15....5D}
\begin{equation}
    P_{ji}=h\nu_{ji}N_j(Z^{+r})A_{ji} \ \ (\text{erg}\ \text{cm}^{-3}\ \text{s}^{-1})
\end{equation}
where $N_j(Z^{+r})$ is the number density of the upper level of the ion, and $A_{ji}$ is the Einstein's coefficient for spontaneous emission of the transition.

The line intensity (radiance) emitted by an area on the Sun is 
\begin{equation}
    I=\dfrac{h\nu_{ji}}{4\pi}\int{N_j(Z^{+r})A_{ji}\text{d}l}=\dfrac{1}{4\pi}\int{P_{ji}\text{d}l} \ \ (\text{erg}\ \text{cm}^{-2}\ \text{s}^{-1}\ \text{sr}^{-1})
\end{equation}
where $\text{d}l$ is the unit length along LOS. The emitting area has been implied in the solid angle. If the emitting area is the unit source area $\text{d}A_s$, then $I$ can be regarded as intensity per unit area.

For Sun-as-a-star observations, the flux (irradiance) of the Sun is related to the volume integral of the unit volume emissivity 
\begin{equation}\label{eq: pintegration}
    F=\dfrac{1}{4\pi d^2}\int_V{P_{ji}\text{d}V}
\end{equation}
where $d$ is the Sun-Earth distance. In this work, we assumed the CME is uniform in both density and temperature; therefore, the unit volume emissivity $P$ will be an invariant within the CME structure at a fixed height. Thus, from Eq. \ref{eq: pintegration} and Eq. \ref{eq:Fontenla_E1} it is derived

\begin{equation}\label{eq: conversion}
    F=\dfrac{1}{4\pi d^2}P\int{\text{d}V}=\dfrac{P\cdot V}{4\pi d^2}=\dfrac{\bar{I}\cdot S}{d^2}
\end{equation}

This leads to
\begin{equation}\label{eq: PFIrelation}
    \dfrac{P\cdot V}{4\pi}=i\cdot V=\bar{I}\cdot S,\quad \text{and}\quad F=\dfrac{i\cdot V}{d^2}
\end{equation}

Here $P$ is the unit volume emissivity (an invariant under uniform CME structure assumption at a given height $h$), $\bar{I}$ is the average intensity per unit area obtained from Hinode/EIS observations, and $S$ is the projected area of CME onto the solar disk. We denoted $P(h)/4\pi$ as $i(h)$.

\begin{figure}[htbp!]
\centering
\includegraphics[width=0.8\textwidth]{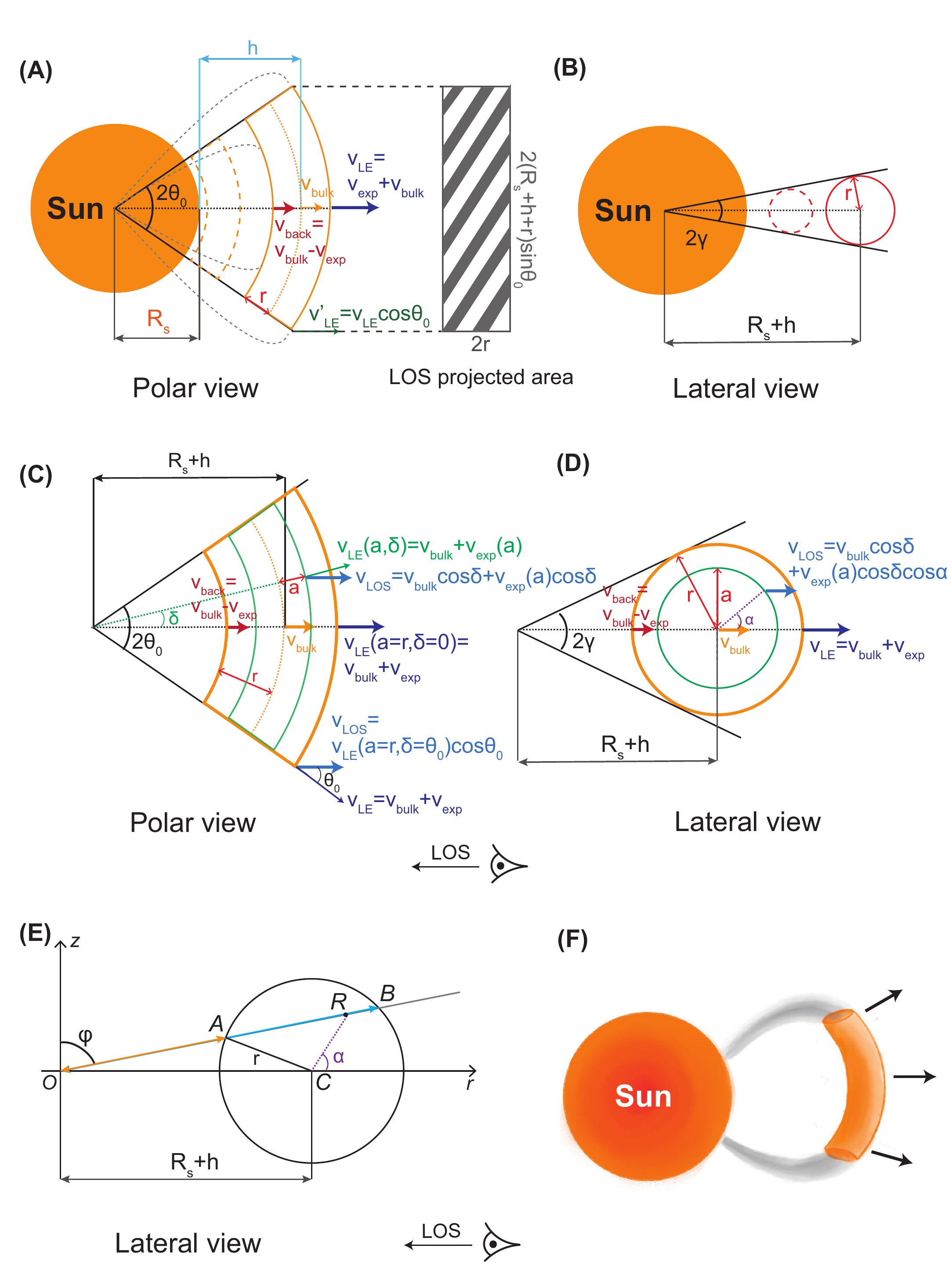}
\caption{The geometric model of the CME structure. The CME is assumed to have a partial donut shape. \textbf{(A)} A self-similarly expanded CME geometry as seen from the polar side. The solid orange curve represents the CME structure (a partial donut shape), and the dashed orange curve depicts the same CME structure a while ago. From the dashed orange shape to solid orange shape, we show the self-similar expansion of the CME structure. The dashed grey curves are the feet of the CME flux tube that we have neglected. The rectangle with grey stripes indicates the LOS projected CME area seen by the observer. \textbf{(B)} The lateral view of CME geometry. The circle represents the circular cross section of the CME flux tube. The transition from the dashed circle to solid circle describes the self-similar expansion. \textbf{(C)} The velocities at different parts of the CME structure (polar view). The orange and green curves indicate different shells within the CME structure. Details can be found in the text. \textbf{(D)} Similar to (C) but for lateral view. \textbf{(E)} The lateral view of the CME structure in a spherical coordinate. This figure is modified from \cite{farmer2005volume}. \textbf{(F)} A cartoon illustration of the CME geometry in this work.} \label{CME_new}
\end{figure}

The average intensity of CME per unit area from solar observation (at the initial height $h_0$ where the observation was made) is denoted as $\bar{I}_{\text{CME}}(h_0)$, and the projected CME area at the initial height is $S_{\text{CME}}(h_0)$. We also assumed that the CME originates from the center of solar disk and propagates along the LOS between the solar disk center and the observer. The LOS projected area can be approximated as a rectangle for the convenience of calculations (our calculations show that the difference between the area of the approximated rectangle and the actual area is less than 1\%). As shown in Figure \ref{CME_new}(A), the approximated LOS projected area at a given height $h$ is 
\begin{equation}
    S_{\text{CME}}(h)=4r\cdot(R_\text{s}+h+r)\sin\theta_0
\end{equation}

From Eq. \ref{eq: PFIrelation} we can write (using the aforementioned relationship $r_0=h_0$)
\begin{equation}
     \bar{I}_{\text{CME}}(h_0)=\dfrac{i(h_0)\cdot V_0}{S_{\text{CME}}(h_0)}=i(h_0)\dfrac{\pi r_0^2(R_\text{s}+h_0)\cdot 2\theta_0}{4r_0\cdot(R_\text{s}+h_0+r_0)\sin\theta_0}
\end{equation}
\begin{equation}\label{eq: lineunitint}
    i(h_0)=\dfrac{2(R_\text{s}+2h_0)\sin\theta_0}{\pi\theta_0\cdot(R_\text{s}+h_0)h_0}\cdot \bar{I}_{\text{CME}}(h_0)
\end{equation}

The emissions of spectral lines are determined by the level populations of the corresponding ions. By solving the statistical equilibrium equations in consideration of all important excitation and de-excitation processes, we are able to calculate the ion level population and identify the most important processes contributing to line emissions \citep{2012uxss.book.....P,2018LRSP...15....5D}. 
Using the procedure \it{pop\_solver.pro}\rm{} in the \sc{chianti}\rm{} software package Ver. 10 \citep{1997A&AS..125..149D,2021ApJ...909...38D}, we found that in the electron density range of $N_\text{e}=10^{8}-10^{10}\ \text{cm}^{-3}$, for the EUV spectral lines, especially Fe \sc{xii}\rm{} 195.12 \AA\ and Fe \sc{xv}\rm{} 284.16 \AA\ used in this work, the upper energy levels of these EUV lines are populated by  collisional excitation. This means the emissivity of these EUV lines is proportional to the square of electron density, which gives $i\propto n_\text{e}^2$ \citep[e.g.,][]{2016JGRA..121.8237L}. Hence we can obtain the height variation of $i(h)$ as 
\begin{equation}\label{eq: lineunitintrat}
    \dfrac{i(h)}{i(h_0)}=(\dfrac{n_\text{e}(h)}{n_\text{e}(h_0)})^2=(\dfrac{R_\text{s}+h}{R_\text{s}+h_0})^{-6}
\end{equation}

\subsubsection{Evolution of Continuum Intensity}\label{continuum}

In addition to spectral line emission, continuum emission is also important when dealing with line profiles. In the EUV wavelength range, free-free emission is the dominant contributor to the continuum intensity \citep{2007A&A...476..675L,2018LRSP...15....5D}. The intensity of free-free emission is dependent on electron density $n_\text{e}$ and ion density $n_\text{i}$ \citep{2018LRSP...15....5D}, using the expression of free-free emission from the aforementioned literature and assuming ionization equilibrium, we can estimate the intensity of free-free emission as $I_{\text{ff}}\propto n_\text{e}^2$. We can then write the equivalent of Eq. \ref{eq: lineunitintrat} for continuum intensity at the two different heights $h_0$ and $h$ as

\begin{equation}
    \dfrac{i_{\text{cont}}(h)}{i_{\text{cont}}(h_0)}=(\dfrac{R_\text{s}+h}{R_\text{s}+h_0})^{-6}
\end{equation}

Similar to Eq. \ref{eq: lineunitint}, based on the observed average continuum intensity $\bar{I}_{\text{cont}}(h_0)$, we have

\begin{equation}
    i_{\text{cont}}(h_0)=\dfrac{2(R_\text{s}+2h_0)\sin\theta_0}{\pi\theta_0\cdot(R_\text{s}+h_0)h_0}\cdot \bar{I}_{\text{cont}}(h_0)
\end{equation}

The flux of continuum is 
\begin{equation}
    F_{\text{cont}}=\dfrac{i_{\text{cont}}\cdot V(h)}{d^2}=\dfrac{\bar{I}_{\text{cont}}\cdot S_{\text{CME}}(h)}{d^2}
\end{equation}

\subsection{Expansion Effect on CME Profiles}\label{expansion}

CME will experience expansion as it propagates. Following previous studies, in our model, the CME structure is assumed to go through self-similar expansion along radial direction. Based on the expansion pattern in previous literature \citep[e.g.,][]{2009CEAB...33..115G,2020SoPh..295..107B}, we have 
\begin{equation}
    v_{\text{LE}}=v_{\text{bulk}}+v_{\text{exp}}
\end{equation}  
\begin{equation}
    v_{\text{LE}}=\frac{\text{d}(R_\text{s}+h+r)}{\text{d}t}
\end{equation}
\begin{equation}
    v_{\text{bulk}}=\frac{\text{d}(R_\text{s}+h)}{\text{d}t}
\end{equation}
\begin{equation}
    v_{\text{exp}}=\frac{\text{d}r}{\text{d}t}=\frac{\text{d}(R_\text{s}+h)}{\text{d}t}\cdot\kappa=\kappa v_{\text{bulk}}
\end{equation}
where $v_{\text{LE}}$ is the speed at the leading edge of the CME, $v_{\text{bulk}}$ is the bulk speed (the centroid speed, which refers to the speed from the velocity evolution profile), $v_{\text{exp}}$ is the expansion speed at the outermost surface of the CME structure (see Figure \ref{CME_new}(A)).

In terms of the self-similar expansion effect, the expansion speed varies from the CME center to the surface. To simulate this effect, we can separate the partial donut structure into many shells with different radii ranging from $0$ to $r$, as shown in Figure \ref{CME_new}(C)-(D). The expansion speed is supposed to increase linearly from the interior to the surface of the flux tube structure. For a shell with radius $a$, the expansion speed at the shell surface is defined as 

\begin{equation}
v_{\text{exp}}(a)=\dfrac{a}{r}v_{\text{exp}}=\kappa\dfrac{a}{r}v_{\text{bulk}}
\end{equation}

In conclusion, the total velocity is $\mathbf{v}=\mathbf{v_{\text{bulk}}}+\mathbf{v_{\text{exp}}}(a)$. Here the bulk velocity and expansion velocity are written in vectors with their directions implied.

\subsubsection{LOS Speed from Different Position Angles}

From spectral observations, we can obtain the LOS projections of CME velocity, therefore, it is important to know the LOS component of velocity in our geometric model. In this work, we only dealt with Earth-facing CMEs, implying that the CME propagates along the direction between the observer and the solar disk center. The LOS velocity in this work refers to the projected velocity along the direction between the disk center and the observer. Considering the expansion effect on the LOS velocity, from Figure \ref{CME_new}(C)-(D), we can see that there are two different position angles that will affect the LOS component of velocity. For polar view, the position angle $\delta$ is defined as the angle between the radial direction and the LOS direction, as shown in Figure \ref{CME_new}(C), and it varies from $-\theta_0$ to $\theta_0$. For lateral view, as revealed in Figure \ref{CME_new}(D), the position angle $\alpha$ is defined as the azimuthal angle on the circular cross section of the flux tube, varying from $-\pi$ to $\pi$ ($\alpha=0$ is defined when the direction is along Earthward direction, and it increases counter-clockwisely.)

Consequently, for a shell with radius $a$, at the position angles $(\delta, \alpha)$, the LOS component of velocity is 

\begin{equation}\label{eq: vlos}
\begin{split}
    v_{\text{LOS}}& =v_{\text{bulk}}\cos\delta+v_{\text{exp}}(a)\cos\delta\cos\alpha\\
    & =v_{\text{bulk}}\cos\delta+\frac{a}{R_\text{s}+h}v_{\text{bulk}}\cos\delta\cos\alpha
\end{split}
\end{equation}

Here $a$ varies from $0$ to $r$, $\delta$ changes between $-\theta_0$ and $\theta_0$， and $\alpha$ falls in the range of $-\pi$ to $\pi$.

\subsubsection{Calculation of Line Profiles in Consideration of Expansion Effect}

At different parts of the CME, self-similar expansion leads to different LOS velocities. These different LOS velocities will broaden the full disk-integrated spectral line profiles. Regarding the CME structure as composed with a large amount of differential elements $\text{d}V$, with each differential element emitting the radiation of $i(h)\cdot \text{d}V$, then the line profile of each element can be obtained based on its LOS velocity, assuming the same line width as that from observations. The overall profiles of the CME should be the summation of profiles over all differential elements. To derive the Sun-as-a-star CME line profile with the inclusion of expansion effect, we will apply a triple calculus in a spherical coordinate.

The following mathematical derivations are adopted and modified from the calculations in \cite{farmer2005volume}, where the volume of a torus was derived in a spherical coordinate $(\rho, \phi, \delta)$. Figure \ref{CME_new}(E) is a schematic illustration showing a simplified lateral view of the flux tube. In this illustration, O is the origin of the spherical coordinate, OC$=R_\text{s}+h$, AC=BC$=r$. In the triplet $(\rho, \phi, \delta)$, $\rho$ refers to OR, $\phi$ is the angle between OR and z as depicted in Figure \ref{CME_new}(E), and $\delta$ refers to the same $\delta$ in Eq. \ref{eq: vlos}. Here $\alpha$ is the same as the position angle $\alpha$ in Figure \ref{CME_new}(D). To derive the line profile, the total line intensity and different LOS velocities for each differential element are required. 

The differential element for triple calculus in a spherical coordinate is written as  
\begin{equation}
    \text{d}V=\rho^2\sin\phi \text{d}\rho \text{d}\phi \text{d}\delta
\end{equation}

From Figure \ref{CME_new}(E) we can see that $\phi$ only varies between a critical angle $\beta$ and $\pi-\beta$, where $\beta=\pi/2-\gamma=\arccos{\kappa}$. For each fixed $\phi$, the variable $\rho$ varies between OA and OB, following \cite{farmer2005volume} we have
\begin{equation}
    \text{OA}=j_1=(R_\text{s}+h)(\sin\phi-\sqrt{\kappa^2-\cos^2\phi})
\end{equation}
\begin{equation}
    \text{OB}=j_2=(R_\text{s}+h)(\sin\phi+\sqrt{\kappa^2-\cos^2\phi})
\end{equation}

We can write the expressions referring to Figure \ref{CME_new}(E)
\begin{equation}
    a=\sqrt{\rho^2+(R_\text{s}+h)^2-2\rho(R_\text{s}+h)\sin\phi}
\end{equation}
and 
\begin{equation}
    \cos\alpha=\dfrac{\rho^2-a^2-(R_\text{s}+h)^2}{2a(R_\text{s}+h)}
\end{equation}

Combining the above expressions with Eq. \ref{eq: vlos}, the LOS velocity for each differential element is
\begin{equation}\label{eq: vcmelos}
    v_{\text{LOS}}(\rho,\phi,\delta)=v_{\text{bulk}}\cos\delta+v_{\text{exp}}(a)\cos\delta\cos\alpha=v_{\text{bulk}}\cos\delta\frac{\rho\sin\phi}{R_\text{s}+h}
\end{equation}

The above equations outline the volume of each differential element, and they also describe how the LOS velocity of each differential element varies as a function of $\rho,\ \phi,\ \delta$. 

For each $\text{d}V$, using Eq. \ref{eq: PFIrelation}, the flux emitted by each differential element is $\text{d}F(h)=\text{d}V\cdot i(h)/d^2$. The variables $i(h)$ and $i_{\text{cont}}(h)$ are constant within the same differential volume. 
Assuming that the line width $\Delta\lambda_D$ of each differential volume remains unchanged during CME propagation, similar to the line profiles constructed from line intensity $\bar{I}$, we can derive the line profiles from the flux of the differential element $\text{d}F$ as 
\begin{equation}\label{eq:fluxprofile}
    f_{\text{d}F}(\lambda)=\text{d}F_\text{p} \exp({-\frac{(\lambda-\lambda_0(1-\dfrac{v_{\text{LOS}}(\rho,\phi,\delta)}{c}))^2}{\Delta\lambda_D^2}})+\text{d}F_\text{cont}
\end{equation}
where $\text{d}F_\text{p}=\text{d}F/(\sqrt{\pi}\Delta\lambda_D)$ and $\text{d}F_\text{cont}=\text{d}V\cdot i_{\text{cont}}(h)/d^2$.

The total flux is calculated as 
\begin{equation}
    F_{\text{CME}}(h)=\dfrac{i(h)\cdot V}{d^2}=\dfrac{i(h)}{d^2}\int_{\delta=-\theta_0}^{\delta=\theta_0}\int_{\phi=\beta}^{\phi=180^{\circ}-\beta}\int_{\rho=j_1}^{\rho=j_2} \text{d}V
\end{equation}

The line profiles of various differential elements of the structure will then be included to generate the full disk-integrated CME line profile, which is the summation of differential line profiles $f_{\text{d}F}(\lambda)$ over the domain of integration:
\begin{equation}\label{eq: fcme}
\begin{split}
    f_{F_{\text{CME}}}(h,\lambda)& =\int{f_{\text{d}F}(\lambda)}\text{d}V\\
    & =\int{\left[\dfrac{i(h)}{\Delta\lambda_D}\exp({-\frac{(\lambda-\lambda_0(1-\dfrac{v_{\text{LOS}}(\rho,\phi,\delta)}{c}))^2}{\Delta\lambda_D^2}})+i_{\text{cont}}(h)\right]\dfrac{\text{d}V}{d^2}}\\
    & =\int{C(h,\lambda)\frac{\text{d}V}{d^2}},
\end{split}
\end{equation}
here we denoted the expression inside the square bracket as $C(h,\lambda)$ for simplicity.

Equations \ref{eq: vcmelos}, \ref{eq:fluxprofile} and \ref{eq: fcme} are the fundamental analytical expressions for Sun-as-a-star spectral line profiles from a propagating CME structure based on our geometric model.

\section{Determination of CME-detectable Instrumental Configurations}
In this section, we will describe the analyzing processes to determine CME-detectable instrumental properties. In Sect. \ref{synthetic}, we introduced the synthetic procedures of Sun-as-a-star line profiles, and the criterion we used to determine the detectability of CME signals from the synthetic profiles. We also performed Monte-Carlo simulations to generate different spectral line profiles and applied the criterion to constrain properties of potential Sun-as-a-star spectrographs in Sect. \ref{monte-carlo}. Moreover, in Sect. \ref{conditions}, we presented the CME-detectable instrumental conditions from our procedures. 

\subsection{Synthetic Sun-as-a-star Line Profiles}\label{synthetic}

In our synthesis, we proposed several different variables of two types to investigate the successful detection of CMEs at different evolution stages from Sun-as-a-star spectral observations. The first type is the instrument parameters, including signal-to-noise ratio (SNR) and spectral resolution of the instrument. The other type is CME-related properties, including $h$, the CME height; $P_{\text{CME}}(h_0)=S_{\text{CME}}(h_0)/S_{\odot}$, the percentage of the initial area of CME region (throughout the paper ``initial'' refers to variables at the height where our EIS observation was made); $P_{\text{AR}^{\prime}}=S_{\text{AR}^{\prime}}/S_{\odot}$, the percentage of the initial area of active region excluding CME area; and $P_{\text{QS}}=1-P_{\text{CME}}(h_0)-P_{\text{AR}^{\prime}}$, the percentage of the initial area of quiet Sun region. For convenience of calculations, we separated the CME regions (also a part of AR) from non-erupted ARs. The percentage of the initial area of the total active region (including CME area) is $P_{\text{AR}}=P_{\text{AR}^{\prime}}+P_{\text{CME}}(h_0)$. We also assumed that, at the initial height (the CME height $h_0$ where EIS observation was made), the area of CME regions should not exceed that of the total active regions. As the CME expands during propagation, the projected area on the solar disk could grow larger than the area of active regions at the initial height. This effect has already been included in our calculations. However, we will only use the initial areas (the ones at the initial height) as our input parameters, avoiding the complexity introduced by increased areas at different CME heights.

\subsubsection{The Overall Sun-as-a-star Spectral Profiles}

At a given height $h$, the overall spectral profile is the full disk-integrated profiles from active regions, quiet Sun regions and CME regions. Considering the CME as an optically thin object in EUV wavelengths, the rest emission (excluding the emission from the CME region) from the solar surface does not vary as the CME propagates. This is also the reason why we separated the area of active regions ($S_{\text{AR}}$) into the one without CME ($S_{\text{AR}^{\prime}}$) and the one where CME height equals the height $h_0$ where it was captured in our EIS observation ($S_{\text{CME}}(h_0)$). However, we should also note that the emission of background active regions can sometimes be enhanced during eruptions (e.g., the increased emission caused by flaring loops). We can write the total Sun-as-a-star flux as 

\begin{equation}
\begin{split}
    F_{\text{tot}}& =F_{\text{CME}}(h)+F_{\text{AR}}+F_{\text{QS}}\\
    & =F_{\text{CME}}(h)+\bar{I}_{\text{AR}}\cdot\dfrac{(S_{\text{CME}}(h_0)+S_{\text{AR}^{\prime}})}{d^2}+\bar{I}_{\text{QS}}\cdot\dfrac{(S_{\odot}-S_{\text{CME}}(h_0)-S_{\text{AR}^{\prime}})}{d^2}
\end{split}
\end{equation}

Equation \ref{eq: PFIrelation} indicates that the flux $F$ and average intensity $\bar{I}$ are related by $F=\bar{I}\cdot S/d^2$, therefore every term in the right-hand-side of the above equation (written in the forms of either $F$ or $\bar{I}$) refers to flux.

Dividing both sides by the solar disk area $S_{\odot}$, we could obtain
\begin{equation}
    \dfrac{F_{\text{tot}}}{S_{\odot}}=\dfrac{F_{\text{CME}}(h)}{S_{\odot}}+\bar{I}_{\text{AR}}\cdot(P_{\text{CME}}(h_0)+P_{\text{AR}^{\prime}})\dfrac{1}{d^2}+\bar{I}_{\text{QS}}\cdot(1-P_{\text{CME}}(h_0)-P_{\text{AR}^{\prime}})\dfrac{1}{d^2}
\end{equation}

The information regarding the CME area is already indicated in the calculated flux $F_{\text{CME}}(h)$, therefore we do not need to know the actual value of the CME area at the height of $h$. Note that $\bar{I}_{\text{QS}}$ are radiances, so the conversion between irradiance (flux) and radiance is applied as indicated in Eq. \ref{eq: conversion}. Similar to Eq. \ref{eq:fluxprofile}, the line profiles created from $\bar{I}_{\text{AR}}$ and $\bar{I}_{\text{QS}}$ are
\begin{equation}
    f_{\bar{I}_{\text{AR}}}=\dfrac{\bar{I}_{\text{AR}}}{\sqrt{\pi}\Delta\lambda_D}\exp({-\frac{(\lambda-\lambda_0)^2}{\Delta\lambda_D^2}})+\bar{I}_\text{cont,AR}
\end{equation}
and
\begin{equation}
    f_{\bar{I}_{\text{QS}}}=\dfrac{\bar{I}_{\text{QS}}}{\sqrt{\pi}\Delta\lambda_D}\exp({-\frac{(\lambda-\lambda_0)^2}{\Delta\lambda_D^2}})+\bar{I}_\text{cont,QS}
\end{equation}

The Sun-as-a-star line profile is then
\begin{equation}
    \bar{f}(\lambda)=\dfrac{\int{C(h,\lambda)\text{d}V}}{S_{\odot}\cdot d^2}+f_{\bar{I}_{\text{AR}}}\cdot(P_{\text{CME}}(h_0)+P_{\text{AR}^{\prime}})\dfrac{1}{d^2}+f_{\bar{I}_{\text{QS}}}\cdot(1-P_{\text{CME}}(h_0)-P_{\text{AR}^{\prime}})\dfrac{1}{d^2}
\end{equation}

In the above equation, the Sun-Earth distance $d$ remains constant. Furthermore, in this work, we only cared about the relative intensities instead of their absolute values, therefore it is convenient to set $d=1$ for simplificity. 

It is clear that the percentage of the flux contributed by active regions and quiet Sun regions depends directly on their areas, which are the input variables in our synthesis. The flux of CME is dependent on both the initial area of the CME ($S_{\text{CME}}(h_0)$) and the CME height ($h$). The information of these variables has already been implied during the calculations of $F_{\text{CME}}(h)$.

\subsubsection{The Effect of Spectral Resolution on Synthetic Profiles}

The spectral resolving power $R$ is the key instrumental parameter for any solar spectrographs, which is related to the smallest resolvable wavelength interval through $\delta\lambda_{\text{instr}}=\lambda/R$. Here $\delta\lambda_{\text{instr}}$ is the spectral resolution in the unit of \AA\ and $\lambda$ is the working wavelength. Since our synthesizd spectral line profiles are based on the analytical expression and the observed spectra from Hinode/EIS, the best spectral resolution we can have is the one of EIS. Therefore, only spectral resolutions no better than that of EIS are investigated in our analysis. We then degraded the spectral resolution based on $\delta\lambda_{\text{instr}}$ of EIS by convolving point spread functions (PSFs) with different widths (in the unit of \AA). This will increase the instrumental widths, leading to worsened spectral resolutions. The generated $\delta\lambda_{\text{instr}}$ is 
\begin{equation}
    \delta\lambda_{\text{instr}}=\sqrt{\delta\lambda_{\text{PSF}}^2+\delta\lambda_{\text{EIS}}^2}
\end{equation}
where $\delta\lambda_{\text{PSF}}$ is the full-width-at-half-maximum (FWHM) of the convolved PSF, and $\delta\lambda_{\text{EIS}}$ is defined as $\delta\lambda_{\text{EIS}}=195.12/R_{\text{EIS}}$ \AA, where $R_{\text{EIS}}$ is $\sim 3936$ \citep{2007SoPh..243...19C}. For the other 284.16 \AA\ line, $\delta\lambda_{\text{EIS}}=284.16/R_{\text{EIS}}$ \AA.

\begin{figure}[htbp!]
\centering
\includegraphics[width=0.8\textwidth]{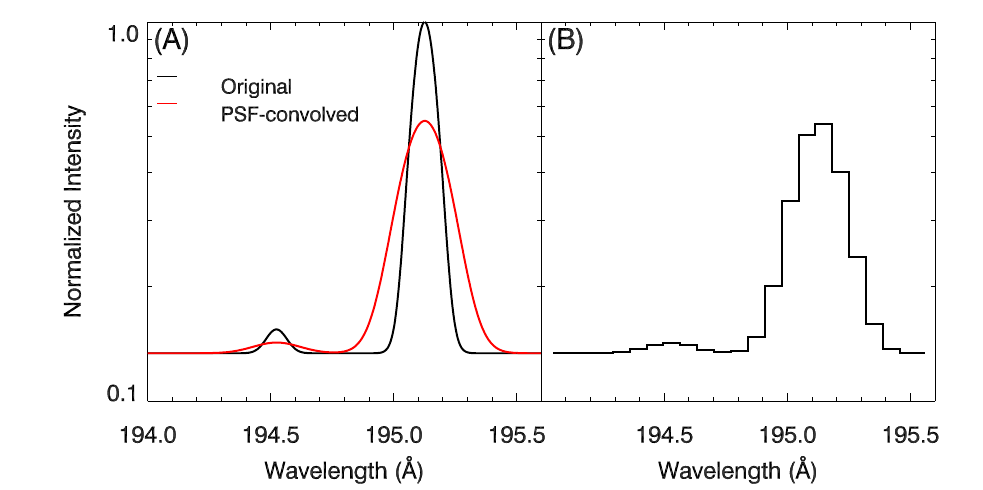}
\caption{The procedures to synthesize full disk-integrated line profiles by considering the effect of spectral resolution. The profiles correspond to the CME height of $h_9=55.30$ Mm. The initial area of CME covers 1\% of the solar disk. And 5\% of the solar disk are active regions (including initial CME area). \textbf{(A)} The black profile is the synthetic profile oversampled to a very small spectral pixel size without any noise nor degradation of spectral resolution. The line profile is scaled with its maximum intensity normalized to unity. The red profile is the PSF-convolved profile with an $\delta\lambda_{\text{instr}}$ of about 0.162 \AA\ (FWHM). It can be seen that the profile is broadened due to increased instrumental width (decreased spectral resolution). \textbf{(B)} The profile sampled with a spectral pixel size of about 0.162 \AA/3.}  \label{fig :psf_instr}
\end{figure}

Due to the increased instrumental widths, a line profile will be broadened as the spectral resolution degrades. We first interpolated the synthetic Sun-as-a-star line profile (without any degradation of spectral resolution) to a very small spectral pixel size, meaning that we were oversampling the line profile. Since the profile was generated from EIS observation, it remained the $\delta\lambda_{\text{instr}}$ of EIS ($\delta\lambda_{\text{EIS}}$), shown as the black line profile in Figure \ref{fig :psf_instr}(A). As we convolved Gaussian PSFs with different FWHMs, we can obtain different broadened line profiles with various $\delta\lambda_{\text{instr}}$ (an example is shown as the red profile in Figure \ref{fig :psf_instr}(A)). As of now, the line profile was the broadened profile due to degradation of spectral resolution, but it was oversampled with a very small spectral pixel size. In order to simulate the regular sampling of spectrographs, we included the degradation of spectral pixel size with degraded spectral resolution. The line profile was then sampled to the corresponding spectral pixel size using IDL function \textit{congrid}. One example is demonstrated in Figure \ref{fig :psf_instr}(B). Here the spectral pixel size is defined as 1/3 of $\delta\lambda_{\text{instr}}$, slightly higher than the Nyquist limit.

\subsubsection{The Effect of SNR on Synthetic Profiles}

To account for the impact of SNR on the observed line profiles, following Equation (3) in \cite{2011ApJ...732...84M}, we added photon noise to each spectral position on the line profile. The line profile with photon noise is represented as \citep{2011ApJ...732...84M}
\begin{equation}
    f(\lambda)=f_{\text{tot}}(\lambda)+\dfrac{\sqrt{f_{\text{tot}}(\lambda)}}{\text{SNR}}\cdot r(\lambda)
\end{equation}
Here $f_{\text{tot}}(\lambda)$ is the synthetic profile with its maximum normalized to unity. The effects of spectral resolution degradation and randomly added photon noise are included. SNR is the signal-to-noise ratio and $r(\lambda)$ is a random number added to each spectral position, which is chosen from a Gaussian distribution with $\mu=0$ and $\sigma=1$.

\subsubsection{The 3-sigma Criterion} \label{sec:3sigma}

\begin{figure}[htbp!]
\centering
\includegraphics[width=0.8\textwidth]{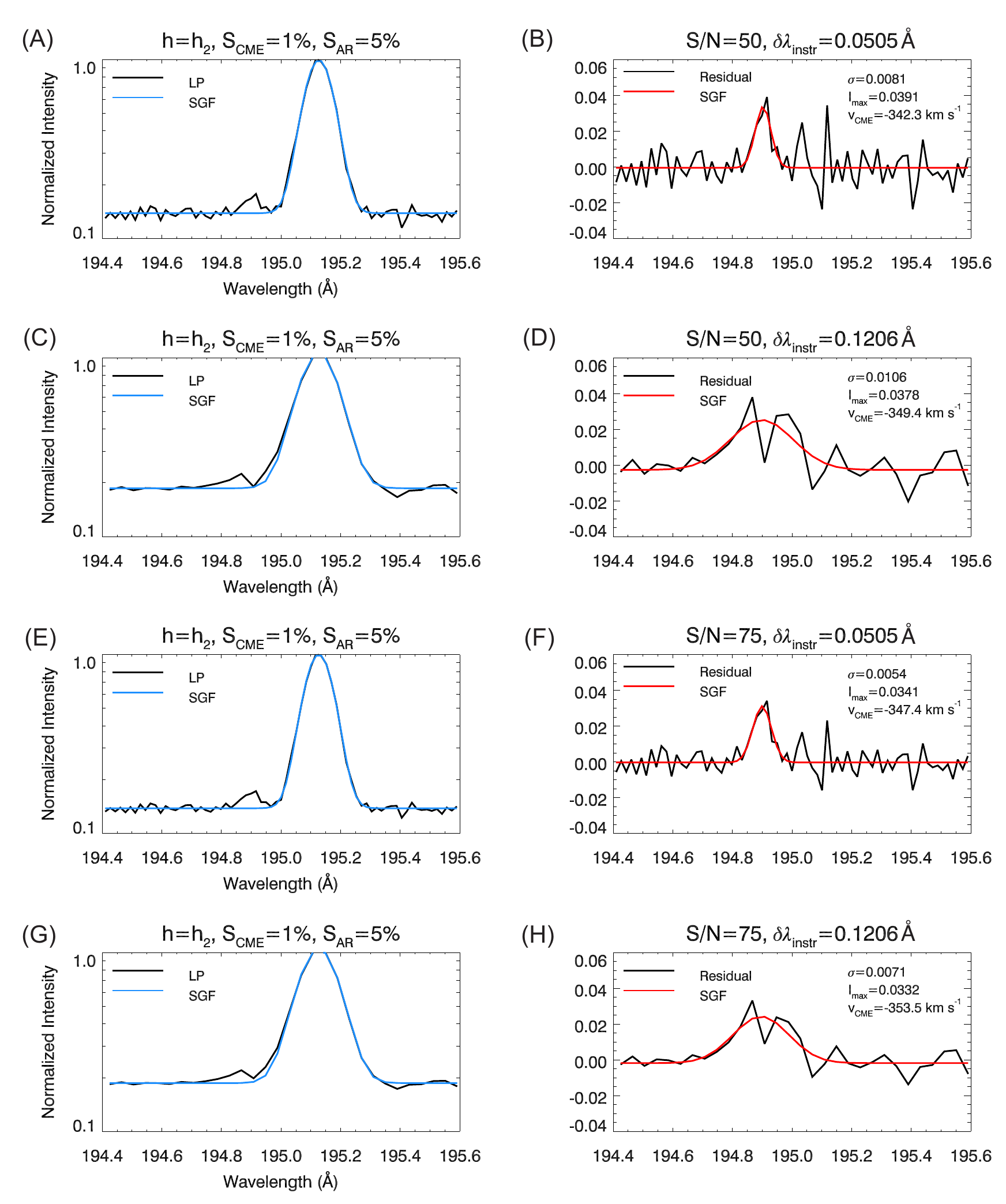}
\caption{Synthetic profiles with different spectral resolutions and SNRs at $h_2=34.20$ Mm with a CME velocity of $v_2=363.50$ km s$^{-1}$. $S_{\text{CME}}=1\%,\ S_{\text{AR}}=5\%$. Here $S_{\text{CME}}$ and $S_{\text{AR}}$ are the initial areas at the height $h_0$.} The left column shows the original profiles (black) and single-Gaussian fitted component (blue), the line profiles were scaled with its maximum intensity normalized to unity The right column shows signals from the CME (residual profile) by subtracting the single-Gaussian fitted components from the original profiles; the red line is the single-Guassian fit to the residual profile. The corresponding SNR and $\delta\lambda_{\text{instr}}$ are denoted in the right column. The calculated $\sigma,\ I_{\text{max}}$ and $v_{\text{CME}}$ are also indicated. Note that we normalized the spectral profiles before adding any noises, so the peak values of the synthetic profiles may sometimes exceed unity after the inclusion of noises (e.g., panels C and G).  \label{fig :profile_init}
\end{figure}

\begin{figure}[htbp!]
\centering
\includegraphics[width=0.8\textwidth]{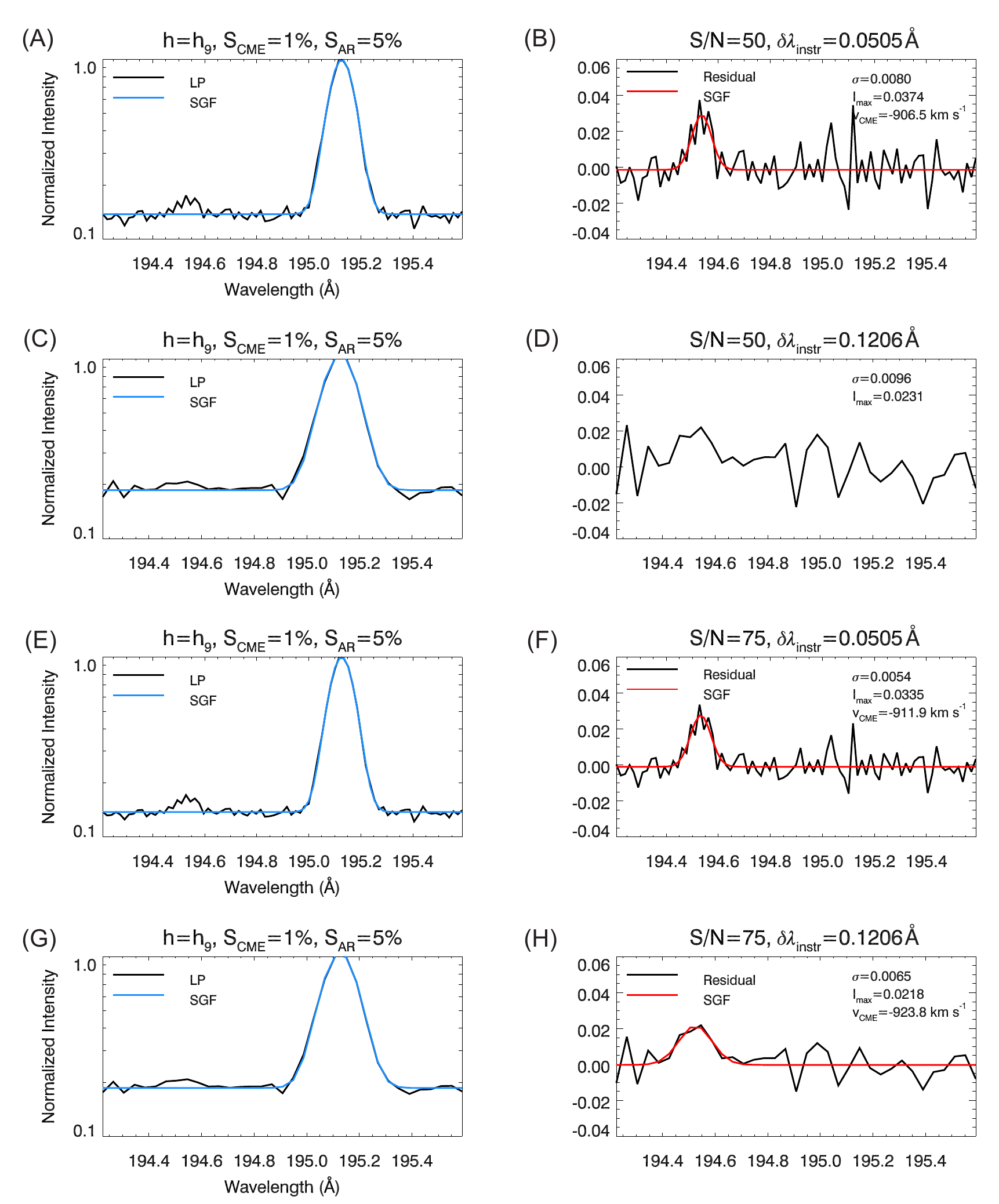}
\caption{Similar to Figure \ref{fig :profile_init}, but for a higher height $h_9=55.30$ Mm with a larger CME velocity of $v_9=930.56$ km s$^{-1}$. $S_{\text{CME}}=1\%,\ S_{\text{AR}}=5\%$. Here $S_{\text{CME}}$ and $S_{\text{AR}}$ refer to the initial areas at the height $h_0$. Note that in panel (D) we did not perform single Gaussian fit for the residual profile since it does not satisfy the 3$\sigma$ criterion.}  \label{fig :profile_esc}
\end{figure}

To determine whether CME signals can be detected from the synthetic line profiles, we used an approach based on a 3-sigma criterion. We first applied a single Gaussian fit to the primary part of the line profile that is barely impacted by CME signals. This allowed only the primary peak of the line profile to be fitted with a single Gaussian function. After that, the fitted single-Gaussian component was subtracted from the ``observed'' profile (the synthetic one with added photon noise and degraded spectral resolution), leaving the contribution from the CME signal (the residual profile). Figure \ref{fig :profile_init} and \ref{fig :profile_esc} show some examples for different variables $S_{\text{CME}},\ S_{\text{AR}}$, SNR and $\delta\lambda_{\text{instr}}$ at different heights (note that throughout the document $S_{\text{AR}}$ refers to the initial area of all active regions including CME, meaning that $S_{\text{AR}}=S_{\text{AR}^{\prime}}+S_{\text{CME}}(h_0)$, thereafter $P_{\text{AR}}=P_{\text{AR}^{\prime}}+P_{\text{CME}}(h_0)$; and $S_{\text{CME}}$ refers to $S_{\text{CME}}(h_0)$).

For the 3-sigma criterion, we defined the $\sigma$ as the standard deviation of the background residual continuum. If the peak intensity of the residual is greater than 3$\sigma$, then we classified the synthetic profiles at the corresponding conditions as ``CME-detectable''. The right columns of Figures \ref{fig :profile_init} and \ref{fig :profile_esc} specify the calculated standard deviation $\sigma$ and the peak intensity of the residual profile $I_{\text{max}}$.

Moreover, if the residual profile satisfies the 3-sigma criterion, we will also compute the CME velocity and the peak intensity of the CME component through a single Gaussian fit to the residual, as shown in red profiles in the right columns of Figures \ref{fig :profile_init} and \ref{fig :profile_esc}. Then we can compare the fitted results with input values as evaluations of accuracy. We will discuss this further in the following parts.

\subsection{Monte-Carlo Simulation}\label{monte-carlo}

\subsubsection{Detectability Based on the 3-sigma Criterion}

\begin{figure}[htbp!]
\centering
\includegraphics[width=0.95\textwidth]{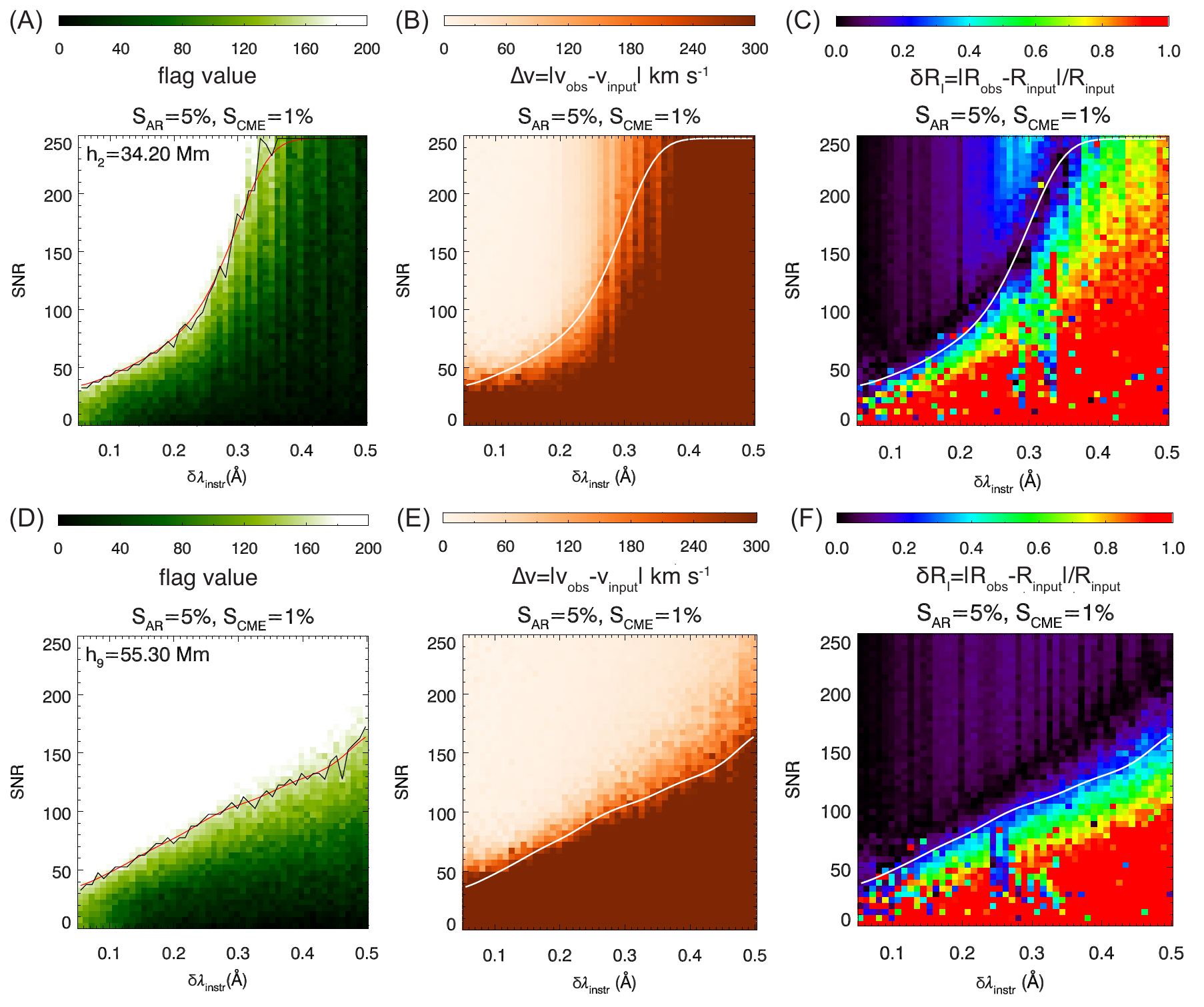}
\caption{Determinations of CME-detectable instrumental parameters using Fe \sc{xii}\rm{} 195.12 \AA. The two rows are two examples based on the synthetic line profiles at different CME heights ($h_2=34.20$ Mm and $h_9=55.30$ Mm). The first column (panels (A) and (D)) are the ``\textit{flag}'' maps under solar activity level of $S_{\text{CME}}=1\%\ \text{and}\ S_{\text{AR}}=5\%$. The color shades represent different values ranging from 0 (dark) to 200 (white). The black solid lines mark the critical line. For SNR and $\delta\lambda_{\text{instr}}$ values above the critical line, over 80\% of the generations of line profiles satisfies the 3-sigma criterion. The red curves are smoothed curves of the black lines, which are used as the critical curves of CME detectability. The second column (panels (B) and (E)) are $\Delta v$ maps related to the accuracy of obtained CME velocity with $S_{\text{CME}}=1\%\ \text{and}\ S_{\text{AR}}=5\%$. The color shades represent $\Delta v$ values. The white curves are the smoothed critical curves determined in (A) and (D). Above the critical curves $\Delta v$ generally can be accurately recovered. The third column (panels (C) and (F)) are the $\delta R_I$ maps related to the accuracy of obtained CME peak intensity from the line profiles with $S_{\text{CME}}=1\%\ \text{and}\ S_{\text{AR}}=5\%$. The different colors represent the values of $\delta R_I=|R_{\text{obs}}-R_{\text{input}}|/R_{\text{input}}$. White curves are the smoothed critical curves. Generally the $\delta R_I$ values above the critical curves are small, meaning that the relative intensity can be accurately derived. The $S_{\text{CME}}$ and $S_{\text{AR}}$ here are the initial areas at the height $h_0$.} \label{fig:f7_sigma_int_vel_map}
\end{figure}

Following the aforementioned procedures, for each fixed $S_{\text{CME}},\ S_{\text{AR}}$, SNR, $\delta\lambda_{\text{instr}}$ and $h$, we performed Monte-Carlo simulations by generating 200 different line profiles with random noise. For each generation, we can define a value \textit{flag} as 1 or 0 according to the 3-sigma criterion. If the 3-sigma criterion is satisfied then \textit{flag}=1, otherwise \textit{flag}=0. For each set of ($h$, $S_{\text{CME}}$, $S_{\text{AR}}$), we summed the $flag$ values over 200 generations, and obtained a two-dimensional map. The value of each pixel on the obtained maps ranges between 0 and 200. Figure \ref{fig:f7_sigma_int_vel_map}(A) and (D) gives two examples of the maps at two different heights ($h_2=34.20$ Mm and $h_9=55.30$ Mm). The abscissa and ordinate of the maps are 
$\delta\lambda_{\text{instr}}$ and SNR, respectively. It is found that for each $\delta\lambda_{\text{instr}}$, the value of each pixel basically increases monotonically with increased SNR. We can then find a critical point where the value reaches 160, which corresponds to a set of \text{SNR} and $\delta\lambda_{\text{instr}}$ values. This means that for this set of instrumental properties, over the 200 generations of line profiles during Monte-Carlo simulations, 80\% of them were identified as CME-detectable, that is, satisfying the 3-sigma criterion. As a result, we could refer to the corresponding \text{SNR} and $\delta\lambda_{\text{instr}}$ as the lower limits of the instrumental parameters to detect CME signals with sufficient accuracy. The critical points were further be connected by a solid black line (critical line) as shown in Figure \ref{fig:f7_sigma_int_vel_map}(A) and (D). This critical line provided the minimum requirements for CME-detetable instrumental parameters at the given $h, S_{\text{CME}}$ and $S_{\text{AR}}$. Due to the increased spectral pixel size with degradation of spectral resolution, sometimes the synthetic line profiles do not have sufficient sampling points (spectral positions of the profile) for single Gaussian fitting to the primary peak. In consequence, if the spectral resolution was too low, uncertainties in the obtained residual profile may arise, and will affect the accuracy on determined detectability using the 3-sigma criterion. As a result of these uncertainties, the critical line is not smooth. Thereafter, for better illustrations, we used a Gaussian smoothed curve (the red curves in Figure \ref{fig:f7_sigma_int_vel_map}(A) and (D)) as the critical curve. The smoothing did not have significant impact on the overall trend of the critical line. The instrumental properties (SNR and $\delta\lambda_{\text{instr}}$) with values above this curve were classified as CME-detectable.

\subsubsection{Accuracy of CME Velocity and Peak Intensity Calculations}\label{sec:vandiaccuracy}

In the previous sections, we have determined the CME-detectable SNR and $\delta\lambda_{\text{instr}}$ based on the 3-sigma criterion. However, to study CMEs through Sun-as-a-star observations, an instrument should not only be capable to detect the CME signal, but also provide accurate CME plasma parameters, especially the velocity and intensity of the CME. As described in Sect. \ref{sec:3sigma}, for line profiles satisfying the 3-sigma criterion, we will determine the CME velocity and relative peak intensity (the peak intensity of CME component relative to that of the primary component) from the residual profiles. In this section, we will investigate the accuracy of calculated CME velocity and peak intensity. Further constraints will be implemented on the required SNRs and spectral resolutions.

In order to evaluate the accuracy of the calculated CME velocity and peak intensity, we defined two parameters: the first is $\Delta v=|v_{\text{obs}}-v_{\text{input}}|$ representing the difference between the calculated CME velocity ($v_{\text{obs}}$) from single-Gaussian fit and the input CME velocity ($v_{\text{input}}$), and the other is $\Delta R_I=|R_{\text{obs}}-R_{\text{input}}|$ representing the difference between the calculated relative CME peak intensity from synthetic profiles ($R_{\text{obs}}=I_{\text{CME}_{\text{obs}}}/I_{\text{P}_{\text{obs}}}$) and the input relative CME peak intensity ($R_{\text{input}}=I_{\text{CME}_{\text{input}}}/I_{\text{P}_{\text{input}}}$). Here the relative peak intensity is the ratio between peak intensities from CME component ($I_{\text{CME}_{\text{obs}}}$, $I_{\text{CME}_{\text{input}}}$) and primary component ($I_{\text{P}_{\text{obs}}}$, $I_{\text{P}_{\text{input}}}$). We can further define two parameters $\delta R_v=\left|(v_{\text{obs}}-v_{\text{input}})/v_{\text{input}}\right|=\left|\Delta v/v_{\text{input}}\right|$ and $\delta R_I=\left|(R_{\text{obs}}-R_{\text{input}})/R_{\text{input}}\right|=\left|\Delta R_I/R_{\text{input}}\right|$, describing the relative deviations of $v_{\text{obs}}$ and $R_{\text{obs}}$.

Similar to Figure \ref{fig:f7_sigma_int_vel_map}(A) and (D), we can plot the 2D images of $\Delta v$ and $\delta R_I$ shown as examples in Figure \ref{fig:f7_sigma_int_vel_map}(B), (C), (E) and (F). We also overplotted the critical curves on the maps. It is to be noted that we only calculated the CME velocity and peak intensity for those line profiles satisfying the 3-sigma criterion. If the profiles did not meet the 3-sigma requirement, then $\Delta v$ is set to be the absolute value of the input CME velocity, and the observed ratio $R_{\text{obs}}$ was set to be two times of the input ratio $R_{\text{input}}$ (thus $\Delta R_I=R_{\text{input}}$). There are some strange behaviors in these $\delta R_I$ maps, but we claim that they do not affect the results of this study. A detailed discussion on these anomalies is provided in Appendix \ref{ap:anomaly}.

\begin{figure}[htbp!]
\centering
\includegraphics[width=0.8\textwidth]{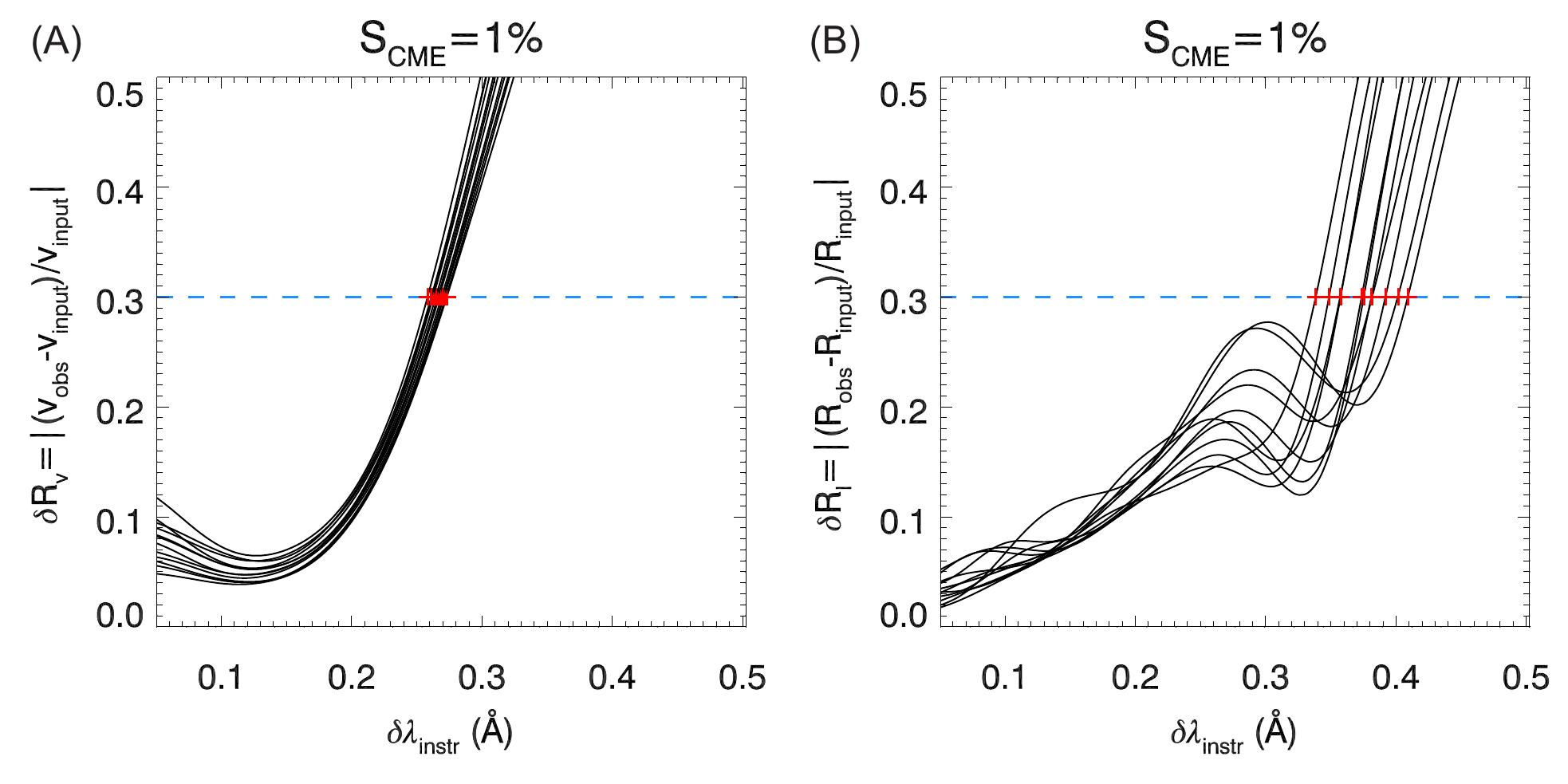}
\caption{The averaged $\delta R_v$ (panel A) and $\delta R_I$ (panel B) values above the critical curves as functions of $\delta\lambda_{\text{instr}}$ at $h_2=34.20$ Mm with $S_{\text{CME}}=1\%$ using Fe \sc{xii}\rm{} 195.12 \AA\ line. Different black curves correspond to different $S_{\text{AR}}$ (ranging from 2\% to 13\%). The blue horizontal dashed lines represent the 30\% level. The red crosses mark the points where $\delta R_v$ and $\delta R_I$ reach this level. The curves are Gaussian smoothed. $S_{\text{CME}}$ and $S_{\text{AR}}$ refer to the initial areas at the height $h_0$.}  \label{fig:int_vel_curve}
\end{figure}

It can be found that above the critical curves, the obtained $\Delta v$ and $\Delta R_I$ (as well as $\delta R_I$) barely change with increased SNR. Instead, they can be treated as a function of only $\delta\lambda_{\text{instr}}$. Based on this, we averaged the $\Delta v$ and $\Delta R_I$ values above the critical curves over the ordinate dimension. This process will generate different $\Delta v$ and $\Delta R_I$ curves as well as $\delta R_v$ and $\delta R_I$ curves as functions of $\delta\lambda_{\text{instr}}$. Figure \ref{fig:int_vel_curve} shows two examples of the curves. Here $\delta R_v$ and $\delta R_I$ estimate the deviations of calculated CME velocity and peak intensity from the input values. From the curves in Figure \ref{fig:int_vel_curve}, we found that the averaged $\delta R_v$ and $\delta R_I$ generally increase with larger $\delta\lambda_{\text{instr}}$ (worse spectral resolution). We can then define a 30\% level of deviation, marked by blue horizontal dashed lines in Figure \ref{fig:int_vel_curve}. For each set of $S_{\text{CME}}$ and $S_{\text{AR}}$, the 30\% level can determine the critical $\delta\lambda_{\text{instr}}$ values. For $\delta\lambda_{\text{instr}}$ larger than this critical value, the obtained CME velocity and peak intensity from the synthetic profiles will have deviations of over 30\% from the input ones. Only instruments with $\delta\lambda_{\text{instr}}$ smaller than the critical values will be considered to be capable for accurate determinations of CME parameters (velocity and relative intensity). We chose the critical $\delta\lambda_{\text{instr}}$ as the smallest $\delta\lambda_{\text{instr}}$ determined from both $\delta R_v$ and $\delta R_I$ curves. The critical $\delta\lambda_{\text{instr}}$ values were then added to previously obtained critical curves. These critical $\delta\lambda_{\text{instr}}$ further constrains the required instrumental spectral resolution to realize both successful detections of CME signals and determinations of CME velocity and peak intensity through line profile asymmetries. Note that the curves were all Gaussian smoothed for better identification of critical points.

\subsection{CME-detectable Conditions} \label{conditions}

\begin{figure}[htbp!]
\centering
\includegraphics[width=0.95\textwidth]{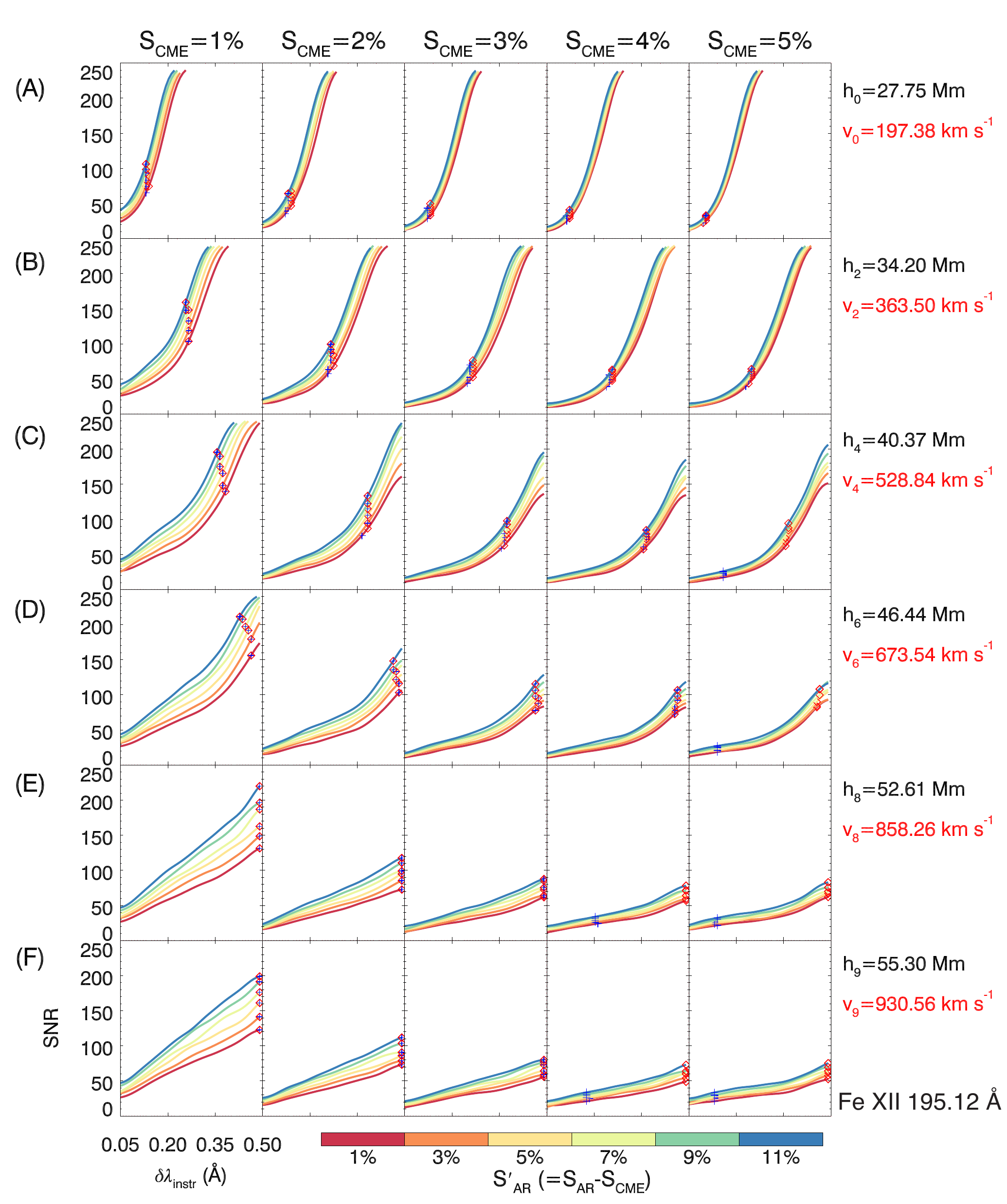}
\caption{The overall constraints on instrumental parameters to detect solar CMEs through line asymmetries using Sun-as-a-star spectral observations. The result is for the Fe \sc{xii}\rm{} 195.12 \AA\ line. Rows (A)-(F) correspond to different heights during the propagation of the CME, the velocity and height are denoted in each row. Each column shows the results for different initial CME areas $S_{\text{CME}}=S_{\text{CME}}(h_0)$. The colored curves in each small panel correspond to different initial $S_{\text{AR}}$ at the height of $h_0$} (e.g., from $S_{\text{AR}}-S_{\text{CME}}=1\%$ (red) to $S_{\text{AR}}-S_{\text{CME}}=11\%$ (blue)), and they are the Gaussian smoothed critical curves determined from the 3-sigma criterion. The blue crosses mark the critical $\delta\lambda_{\text{instr}}$ determined when both CME velocity and peak intensity can be accurately derived, and the red diamonds are those determined when only the accuracy of CME velocity is evaluated.  \label{fig :detect_final_cme_curve}
\end{figure}

Combining the critical curves using 3-sigma criterion (Sect. \ref{sec:3sigma}) with the critical $\delta\lambda_{\text{instr}}$ through evaluations of calculated CME velocity and peak intensity (Sect. \ref{sec:vandiaccuracy}), we obtained the CME-detectable requirements of the instrument, as demonstrated in Figure \ref{fig :detect_final_cme_curve}. In this figure, rows (A)-(F) represent results at different CME heights; the columns are results under different initial CME areas $S_{\text{CME}}$. The figure consists of 36 small panels. In each panel, the abscissa and ordinate are 
$\delta\lambda_{\text{instr}}$ (ranging from $\sim 0.05\ $\AA\ to $\sim 0.5\ $\AA\ for Fe \sc{xii}\rm{} 195.12 \AA) and SNR (from 0 to 200), respectively. The curves with different colors are the smoothed critical curves (using the 3-sigma criterion) for different $S_{\text{AR}}$. There are also blue crosses on each curve, marking the critical $\delta\lambda_{\text{instr}}$ determined when both CME velocity and peak intensity can be determined accurately. Under different conditions (i.e., different heights $h$, different initial CME area $S_{\text{CME}}$ and active region area $S_{\text{AR}}$), for a set of $\delta\lambda_{\text{instr}}$ and SNR lying above the critical curves, if $\delta\lambda_{\text{instr}}$ is also smaller than the critical $\delta\lambda_{\text{instr}}$ marked by the blue crosses in Figure \ref{fig :detect_final_cme_curve}, we will know that there is a 80\% chance that we can confidently identify CME signals from the observed line profiles, and the calculated CME parameters are within acceptable accuracy ranges. Therefore, we can conclude that this set of $\delta\lambda_{\text{instr}}$ and SNR of the instrument is suitable for detections of CMEs through Sun-as-a-star spectral observations. If we only need accurate diagnostics of CME velocity, we can refer to the red diamonds on these curves as the critical $\delta\lambda_{\text{instr}}$. We will show in the following that a precise determination of velocity is easier than that of peak intensity. In a word, this figure includes all the requirements of instrumental parameters to accurately detect CMEs under different conditions.

\begin{figure}[htbp!]
\centering
\includegraphics[width=0.95\textwidth]{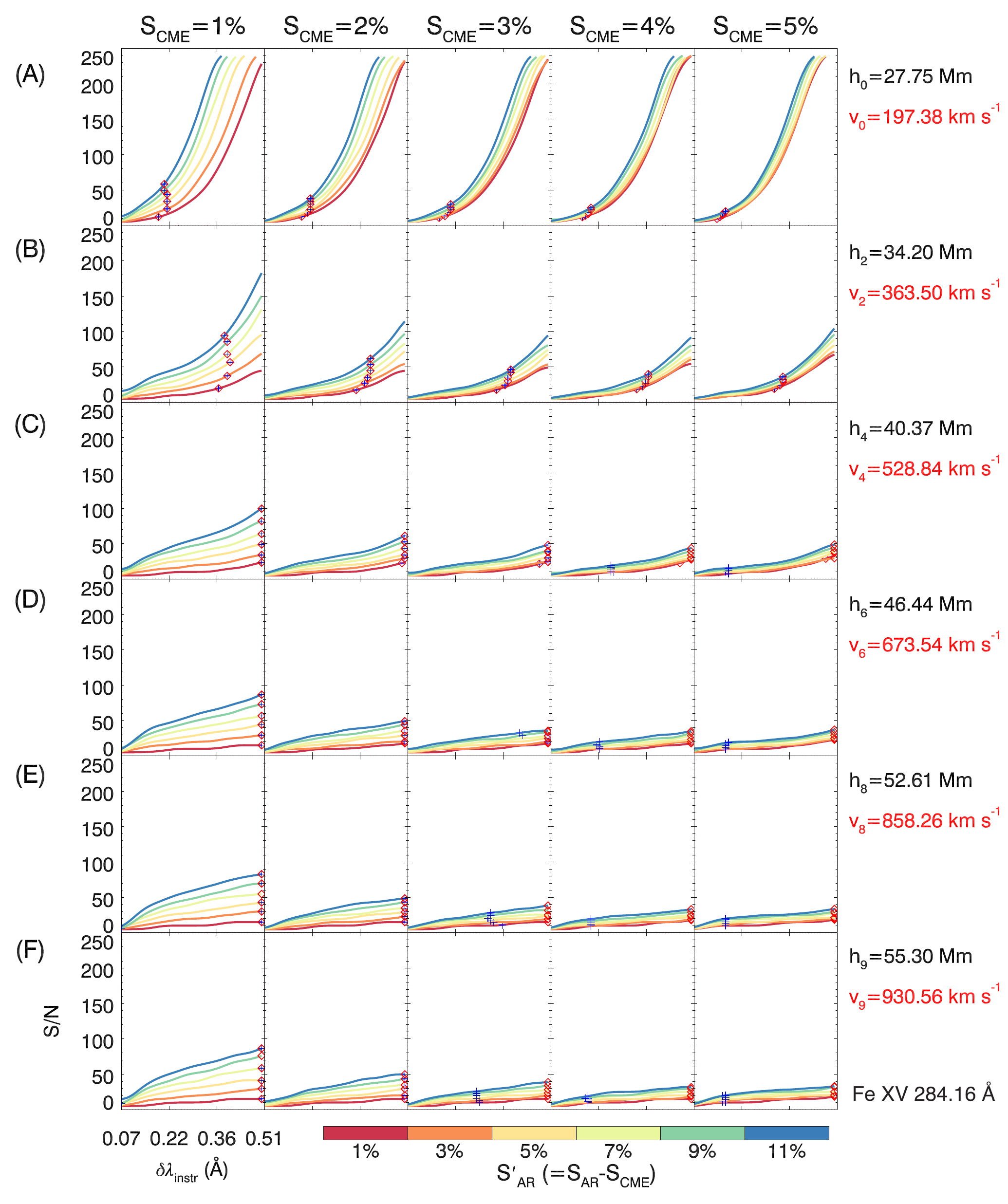}
\caption{Similar to Figure \ref{fig :detect_final_cme_curve} but for the Fe \sc{xv}\rm{} 284.16 \AA\ line. Here $S_{\text{CME}}$ and $S_{\text{AR}}$ are the initial areas at the height $h_0$.}  \label{fig :detect_final_cme_curve_FeXV}
\end{figure}

Similar to the results using Fe \sc{xii}\rm{} 195.12 \AA\ line, we can also do calculations and synthesis for the hotter Fe \sc{xv}\rm{} 284.16 \AA\ line. Figure \ref{fig :detect_final_cme_curve_FeXV} demonstrates the overall constraints on instrumental parameters for Fe \sc{xv}\rm{} 284.16 \AA. The $\delta\lambda_{\text{instr}}$ ranges from $\sim 0.07\ $\AA\ to $\sim 0.5\ $\AA.

\section{Results and Discussion}\label{sec: discuss}
\subsection{Results}

In this work, using the observed solar EUV spectral profiles as references, based on several assumptions and models, we have derived the analytical expression of Sun-as-a-star spectral line profiles during early propagation of a CME.

Utilizing the analytical expressions, we synthesized different Fe \sc{xii}\rm{} 195.12 \AA\ and Fe \sc{xv}\rm{} 284.16 \AA\ line profiles with varied CME conditions (i.e., different heights $h$, different initial CME area $S_{\text{CME}}$ and active region area $S_{\text{AR}}$) and instrument properties (spectral resolution and signal-to-noise ratio). By performing Monte-Carlo simulations and applying a 3-sigma criterion, we first determined CME-detectable instrumental parameters for different CME conditions; moreover, by evaluating the accuracy of calculated CME velocity and peak intensity, we further constrained the required spectral resolution ($\delta\lambda_{\text{instr}}$) for accurate diagnostics of CME parameters.

From Figure \ref{fig :detect_final_cme_curve} (for Fe \sc{xii}\rm{} 195.12 \AA) and Figure \ref{fig :detect_final_cme_curve_FeXV} (for Fe \sc{xv}\rm{} 284.16 \AA) we can see that, at lower heights where the CME velocity is fairly small, the spectral component of the erupted CME plasma can not be completely separated from the primary spectral component. As a result, at these lower heights, spectral resolution plays a major role in the identification of CMEs from Sun-as-a-star spectral observations. As expected, with a worse spectral resolution, the required SNR will increase. When the CME has propagated to a higher height, where its velocity becomes large enough for the two components to be separated in the spectra, the detectability is less dependent on the spectral resolution. Moreover, as the initial area of CME ($S_{\text{CME}}$ or $P_{\text{CME}}$) increases compared to that of active regions, the required SNR will be lower. 

It is important to note that, as the spectral resolution worsens, there will be less available sampling positions for Gaussian fitting, as well as for the application of the 3-sigma criterion, therefore increasing the uncertainty. Under certain circumstances, artificial effects will be introduced to a variety of worsened spectral resolutions, producing discontinuities seen in the \textit{flag} maps, as well as the intensity and velocity evaluation maps in Figure \ref{fig:f7_sigma_int_vel_map}. However, they have little impact on the overall trends of the critical curves we obtained here. We claim that higher uncertainties should be included when considering those worse spectral resolutions in our work.

In our calculations, the geometric model relates the initial area of the CME ($S_{\text{CME}}$ or $P_{\text{CME}}$) to the angular width of it ($2\theta_0$ in Figure \ref{CME_new}(A)). Previous statistical studies suggest that the typical angular widths of CMEs are around $50^{\circ}-60^{\circ}$ \citep{2014GeoRL..41.2673G,2021FrASS...8...73P}, corresponding to initial areas (the area at the height where the CME was captured by EIS in our observation) of a CME covering $1-3\%$ of the solar surface as defined in our analysis. As a result, we regarded $S_{\text{CME}}$ with $1-3\%$ of $S_{\odot}$ as typical CME areas at our initial height $h_0$.

Based on the above conditions, we hereby refer to the second column in our Figures \ref{fig :detect_final_cme_curve} and \ref{fig :detect_final_cme_curve_FeXV}. To detect successfully erupted CME with a velocity exceeding the local escape velocity (e.g., at the height of $h_6=46.44$ Mm with $v_6=673.54\ \text{km}\ \text{s}^{-1}$), for Fe \sc{xii}\rm{} 195.12 \AA, the required minimum SNR is about 50-100. We also require a $\delta\lambda_{\text{instr}}$ better than $\sim 0.4$ \AA\ (a spectral resolving power of $R\sim 500$) so that the velocity and peak intensity of the CME can be accurately derived (see second column of row (D) in Figure \ref{fig :detect_final_cme_curve}). For the hotter Fe \sc{xv}\rm{} 284.16 \AA\ line, since most emission of this line originates from active regions, the instrumental requirements are much looser. The required minimum SNR is about 10-50, and for the $\delta\lambda_{\text{instr}}$ we chose, the CME can always be detected with accurate diagnostics of velocity and intensity (from $0.07$ \AA\ to $0.5$ \AA, corresponding to spectral resolving powers of $R\sim 500-4000$) (see second column of row (D) in Figure \ref{fig :detect_final_cme_curve_FeXV}).

From our analysis, we found that the accurate determination of CME peak intensity is more difficult than that of CME velocity, possibly because of the degradation of spectral resolution, the introduced photon noise, and the fact that the noiseless CME profiles calculated from our model were not in perfect Gaussian shapes. As the spectral resolution becomes worse, the instrumental widths will dominate the line widths of the profile, leading to decreasing peaks of both CME component and primary component. The reduction of the peak intensities obviously raises uncertainties in determined $\delta R_I$. Furthermore, when considering the expansion effect of the CME structure, the obtained CME profiles may be distorted from a Gaussian shape. This effect is most prominent for $S_{\text{CME}}=4-5\%$, leading to a large deviation between the calculated parameters and the input ones. Thereafter the critical $\delta\lambda_{\text{instr}}$ is very small due to large uncertainties on intensity calculations, which is likely to be overstrict (see the blue crosses from panels (D)-(F) in the fourth and fifth columns in Figures \ref{fig :detect_final_cme_curve} and \ref{fig :detect_final_cme_curve_FeXV}). Opposing to the intensity, the velocity can be more accurately determined, because the line broadening and shape distortion did not affect the line centroid too much. Therefore, if we only constrain the instrumental parameters based on the accuracy of velocity diagnostics, the requirements on the instruments could be looser.

We only used a fast-accelerated CME model in our analysis. To do evaluations for more typical CMEs with slowly increasing velocities, we adopted a moderate-accelerated CME kinematic model \citep[H1 in][] {2020ApJ...894...85C} and performed similar analyses. The results are summarized in Appendix \ref{ap:ex}. Similar approaches may also be applied to study the detection of CMEs on other solar-like stars through EUV spectral line asymmetries, with some alterations required. This will be discussed in a future related work. We noticed that X-ray spectroscopic observations have been used to detect stellar CME candidates \citep[e.g.,][]{2019NatAs...3..742A}. In the future we may also reconstruct line profiles during stellar CMEs by adjusting our model parameters such as the selected wavelengths, spectral lines and stellar radius. A comparison through spectral line synthesis with the existing observations may allow us to estimate the relative contributions from quiescent, active and CME regions to the observed profiles during stellar CMEs.

\subsection{Discussions on the Methodology}

In order to derive the analytical expression of the Sun-as-a-star line profiles during CMEs, we adopted several assumptions and models. Here we will discuss the impacts of these assumptions and models on our results.

The line profile of an erupting structure is mainly determined by the velocity of the motion through Doppler effect and the line intensity. 

The velocity of the motion (CME bulk velocity) relies on the kinematic evolution of the CME. We first adopted a best-fitted kinematic model describing the initial and main-acceleration phases of the CME from a previous work. Comparing with the observed velocity from Hinode/EIS, we found the CME should have a very large acceleration rate. As a result, it is accelerated, very soon after the initiation, to the local escape velocity, and can be considered as a successful eruption. Since the CME experiences not only upward bulk motions but also expansion motions, and both motions will greatly impact the shapes of line profiles through Doppler effects, we constructed a geometric model that is both physically reasonable and analytically expressible to describe the CME struture. This model is evolved from the conventional GCS model which describes a flux tube of CME. Combined with the widely used self-similar expansion assumption, we can determine the LOS velocities at different parts of the CME structure. 

We also derived the analytical expression of line intensity as the CME propagates. To do this, following previous literature, we used several assumptions, including the mass conservation, uniform density distribution and isothermal conditions of the CME structure. As mentioned earlier, even though such assumptions, particularly the uniform density distribution and isothermal conditions, are likely to be over-simplifications, for the purpose of deriving an analytical expression, we preferred the simple assumptions that could in principle describe the physical conditions of a CME. In fact, the density and temperature distributions of CMEs are still not fully studied, and they vary from event to event. For example, in some well-developed models such as the GCS model, no temperature information is provided, and the density is highly dependent on the fitting to white-light observations. Therefore, as mentioned in the previous sections, the density and temperature assumptions used in our work are also commonly adopted in other CME studies. Nevertheless, it is important to note that these assumptions still introduce uncertainties to our synthetic results. As a case in point, in our analysis, the assumptions of mass conservation and self-similar expansion led to monotonically decrease of density, as described by Eq. \ref{eq:density}. However, a lot of researches have shown that the mass and the density of CMEs will increase with height at early stage due to the pileup of plasma in the front \citep[e.g.,][]{2013ApJ...768...31B,2015ApJ...812...70F,2018SoPh..293...55H}. In consequence, the increase of CME mass and density yields the enhancement of intensity, as opposed to the decreasing intensity with height derived in our work. This may mean that our synthetic CME intensities are probably underestimated under certain situations, so the required SNR and spectral resolution could be lower than our current results. In addition, multi-temperature structures are also observed in CMEs \citep[e.g.,][]{2011ApJ...732L..25C,2013ApJ...769L..25C}. \cite{2019ApJ...874..164R} constructed the temperature profiles as a function of radial height in different plasma components of CMEs using empirical models. It can be found that during the early propagation of CMEs, the temperature of coronal plasma in the CME barely changes, but the temperatures of plasma in prominence-corona transition region (PCTR) around the CME core will increase rapidly from low temperatures of $\sim 10^4-10^5 K$ to coronal temperature. The EUV lines we used in this work are formed at coronal temperatures. This means that at the beginning of the eruption, most of the emissions of the two spectral lines originate from coronal plasma surrounding the relatively cool CME core. However, as the CME propagates, the heated PCTR plamsa may also contribute to the emissions of the two coronal lines. In this case, the line intensity as derived under the isothermal assumption of CME structures in our analysis will deviate from the intensity contributed by multi-thermal plasma components of CMEs. This effect of isothermal assumption on our current results requires further investigations in the future.

During the calculations, we also assumed that the intensity of EUV lines is proportional to the square of electron density ($I\propto n_e^2$). This is generally a reasonable assumption. For coronal approximations \citep[low density, see][]{2018LRSP...15....5D} where the ionization fraction of the element does not depend on electron density, the intensity $I$ is purely proportional to $n_\text{e}^2$ \citep{2020MNRAS.491..576J}. But for precise calculations when the coronal approximation is not used, the dependence of $I$ on $n_\text{e}^2$ varies with different lines. As can be calculated using \it{pop\_plot.pro} \rm{} in \sc{chianti}\rm{}, for Fe \sc{xv}\rm{} 284.16 \AA, its intensity is almost proportional to $n_\text{e}^2$ for electron densities varying between $10^8-10^{10}\ \text{cm}^{-3}$. However, for Fe \sc{xii}\rm{} 195.12 \AA, the line intensity will be more sensitive to $n_\text{e}^2$ at lower densities. This means that as the CME propagates and its density decreases, the emissions of Fe \sc{xii}\rm{} 195.12 \AA\ from the CME will be more prominent than we predicted under current assumptions. It will be easier to detect CME signals through line asymmetries. Under these circumstances, even with smaller SNRs and worse spectral resolutions compared to our current results, the signal of CMEs could still be detected.

We did not include the influences of adjacent spectral lines in the selected wavelength ranges. However, in the blue wing of Fe \sc{xii}\rm{} 195.12 \AA, there are two lines, one is Fe \sc{viii}\rm{} 194.66 \AA\ line \citep[e.g.,][]{2007PASJ...59S.857Y} and the other is an unidentified line near 194.8 \AA\ which has been observed in several previous works \citep[e.g.,][]{2008ApJS..176..511B,2012A&A...537A..38D}. Although quite faint compared to their neighbor with strong emissions, they could still be blended with the blue-shifted CME component if the CME velocity is high enough (e.g., $\sim 500\ \text{km}\ \text{s}^{-1}$). This may affect our assessment. For the wavelength range of Fe \sc{xv}\rm{} 284.16 \AA,  a cooler Al \sc{ix}\rm{} 284.06 \AA\ line that is close to the target line centroid is present. Even though it is very weak in active regions, this Al \sc{ix}\rm{} 284.06 \AA\ line will have almost equivalent emission as the hotter Fe \sc{xv}\rm{} 284.16 \AA\ line in quiet Sun regions \citep{2007PASJ...59S.857Y}. Nevertheless, due to the small wavelength difference between the two lines, the analysis will only be possibly affected for CMEs with very small velocities (for example $\sim 100\ \text{km}\ \text{s}^{-1}$). For the successfully erupted situations (exceeding local escape velocity), we claim that the results will hardly be affected. 

In our work, we did not consider changes in line profiles caused by flare-origin flows. During flares, the chromospheric plasma could be rapidly heated by the accelerated electrons, resulting in plasma flows with enhanced emission which are referred to as chromospheric evaporation. According to the energy flux of the electrons, chromospheric evaporation can be categorized as “gentle evaporation” and “explosive evaporation” \citep[e.g.,][]{2006ApJ...638L.117M,2009ApJ...699..968M,2015ApJ...811..139T,2019ApJ...879...30L}. For gentle evaporation, the evaporation flows have small velocities of a few to a few tens of $\text{km}\ \text{s}^{-1}$. For explosive evaporation, plasma formed near the formation temperature of Fe \sc{xii}\rm{} normally moves downwards, thus showing redshifts, which will hardly impact the blue-shifted profile caused by CMEs \citep[e.g.,][]{2009ApJ...699..968M}. The plasma formed around the formation temperature of Fe \sc{xv}\rm{} has a small upward velocity of around $30\ \text{km}\ \text{s}^{-1}$ \citep[e.g.,][]{2009ApJ...699..968M}. Since only CMEs with velocities exceeding $\sim 100\ \text{km}\ \text{s}^{-1}$ (e.g., Figure \ref{fig :detect_final_cme_curve}(A) and Figure \ref{fig :detect_final_cme_curve_H1}(A)) can be well detected, the small blueshifts associated with these flows can barely impact the secondary components contributed by CMEs even during their early propagation phase. However, the intensities of the main components of the line profiles may increase during flares. \cite{2014ApJ...789...61M} analyzed the full disk-integrated intensity enhancements of different spectral lines during a flare from SDO/EVE. They showed that for the two lines we used here, the total intensity will increase by $\sim 2-10\%$. Since the increased intensities are from plasma with a nearly zero or small velocity, only the emissions of the main peak in our synthetic profiles will be sligtly enhanced. The emission of CME components in the Sun-as-a-star profiles will remain the same, thus the required SNR will not change. Therefore, we claim that under the SNR and spectral resolution conditions described in our work, the detectability of CME signals from full disk-integrated spectral line profiles is likely not significantly impacted by flare-related velocity components and emission enhancements.

\section{Summary}

Accurate predictions of CME trajectory and arriving time at the Earth rely on the precise measurements of its speeds and propagation directions. A low-cost EUV spectrograph that can perform Sun-as-a-star observations with high cadence could be important for the spectral diagnostics of solar CMEs, including reconstruction of CME velocity vector.

In this study, to investigate the possibility of CME detections through EUV line profile asymmetries using Sun-as-a-star spectrographs, we synthesized different full disk-integrated line profiles during CMEs under various conditions. Based on solar spectral observations, models and assumptions, we derived the analytical expressions for Sun-as-a-star line profiles of CMEs. These analytical expressions can be used to synthesize full disk-integrated line profiles for both solar and stellar CMEs. For different solar conditions (the CME height $h$ and solar activity related variables such as the areas of CME regions $S_{\text{CME}}$ and active regions $S_{\text{AR}}$) and instrumental configurations (spectral resolution and signal-to-noise ratio), the synthetic line profiles were tested by a 3-sigma criterion, to determine the minimum SNR and spectral resolution required to detect CME signals. We further evaluated the accuracy of CME velocity and intensity computed from the synthetic line profiles. 
We found that for typical solar conditions (the areas of CME and ARs at the ``initial'' height of $h_0\approx 27.75\ $Mm are around $1-3\%$ and less than $\sim$10\% of the solar surface, respectively), using Fe \sc{xii}\rm{} 195.12 \AA\ line, an SNR of 50-100 is required with a minimum spectral resolving power of $R_{\text{min}}\approx 500$; for the hotter Fe \sc{xv}\rm{} 284.16 \AA\ line, the required SNR is about 10-50 and $R_{\text{min}}\approx 500$.

Our study provides important constraints for the design of future Sun-as-a-star EUV spectrographs that aim to monitor solar CMEs through spectral line asymmetries. Our model and analysis methods can also be used to study the detectability of stellar CMEs using EUV spectra.

\begin{acknowledgments}
This work is supported by NSFC grants 11825301 and 11790304. We would like to thank Drs. Peter Young, Ming Xiong, Li Feng and Guiping Zhou for helpful discussions and comments on the manuscript. \textit{Hinode} is a Japanese mission developed and launched by ISAS/JAXA, with NAOJ as domestic partner and NASA and STFC (UK) as international partners. It is operated by these agencies in co-operation with ESA and NSC (Norway). CHIANTI is a collaborative project involving George Mason University, the University of Michigan (USA), and the University of Cambridge (UK). Y.G. is supported by NSFC (11773016 and 11961131002) and 2020YFC2201201. X.C. is funded by NSFC (11722325 and 11733003) and 2021YFA1600504.

\end{acknowledgments}

\appendix

\section{The Anomaly in $\delta R_{\text{I}}$ Maps} \label{ap:anomaly}

\begin{figure}[htbp!]
\centering
\includegraphics[width=0.99\textwidth]{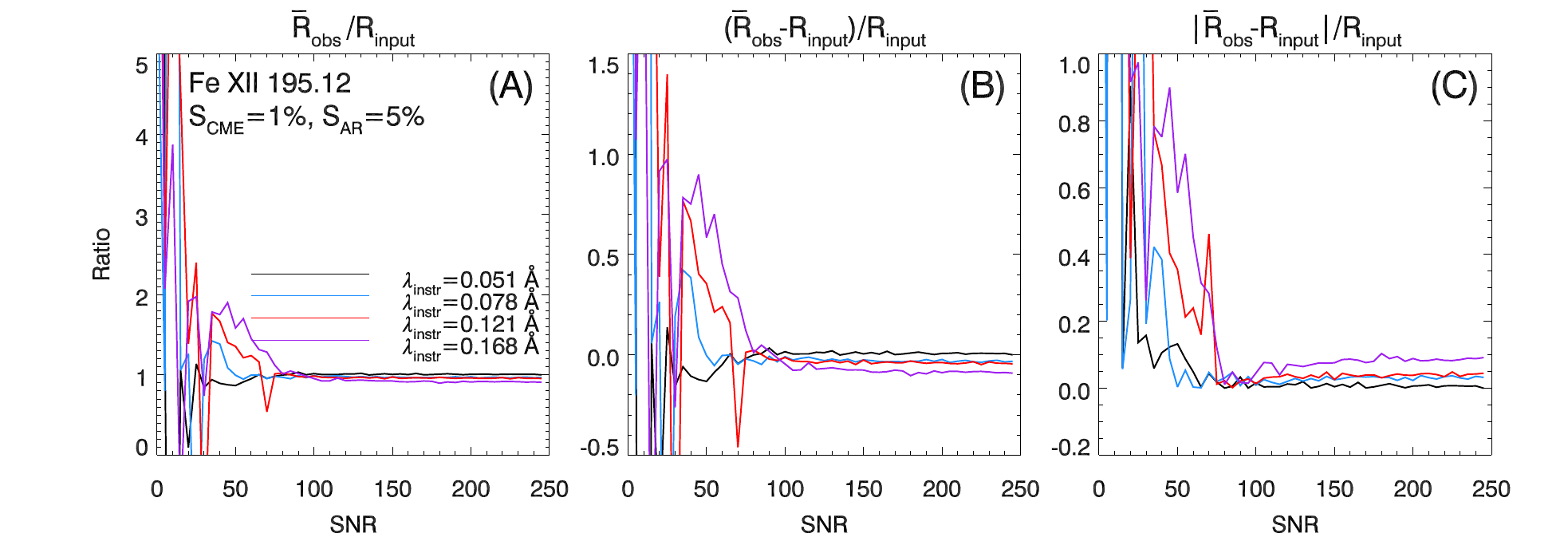}
\caption{The variations of $R_{\text{obs}}$ and the related values for a fixed $\delta\lambda_{\text{instr}}$ as a function of SNR. Different colors represent curves obtained for different $\delta\lambda_{\text{instr}}$. The figures are the results related to Figure \ref{fig:f7_sigma_int_vel_map}(C). \textbf{(A)} For a chosen $\delta\lambda_{\text{instr}}$, we plot the the value $\bar{R}_{\text{obs}}/R_{\text{input}}$ as a function of SNR. It is clear that $\bar{R}_{\text{obs}}/R_{\text{input}}$ will first go through large oscillations due to the very low SNR, then generally decrease with increasing SNR and finally approach a nearly constant value. \textbf{(B)} The value $(\bar{R}_{\text{obs}}-R_{\text{input}})/R_{\text{input}}$ as a function of SNR. This panel shows the same trend as panel (A). \textbf{(C)} $|\bar{R}_{\text{obs}}-R_{\text{input}}|/R_{\text{input}}$ as a function of SNR. Due to the systematic deviations during the calculations, $(\bar{R}_{\text{obs}}-R_{\text{input}})/R_{\text{input}}$ will decrease to a minimum near the critical SNR (the signal-to-noise ratio determined from the critical curve at a given $\delta\lambda_{\text{instr}}$). This will result in a ``reflection'' of the $(\bar{R}_{\text{obs}}-R_{\text{input}})/R_{\text{input}}$-SNR curve, which means that $(\bar{R}_{\text{obs}}-R_{\text{input}})/R_{\text{input}}$ first decreases to a minimum, and then slowly increases to almost a constant. Note that $S_{\text{CME}}$ and $S_{\text{AR}}$ are the initial areas at the height $h_0$.}  \label{fig :anomaly_curve}
\end{figure}

As mentioned in Sect. \ref{sec:vandiaccuracy}, there are some anomalies in the $\delta R_{\text{I}}$ maps in Figure \ref{fig:f7_sigma_int_vel_map}(C) and (F). For example, in Figure \ref{fig:f7_sigma_int_vel_map}(C), the $\delta R_I$ values reaches their minimum at regions near the critical curve (the white curve). This is mainly due to the uncertainty in the peak intensity calculation and the following statistical deviation in Monte-Carlo simulations. As demonstrated in Figures \ref{fig :profile_init} and \ref{fig :profile_esc}, the residual profiles are usually very noisy, leading to large uncertainties during single-Gaussian fitting, as well as on the resulted peak intensity. When generating these $\delta R_I$ maps, we averaged over 200 generations. For the 200 generated line profiles, if they satisfies the 3-sigma criterion, then the peak intensity values were set to be the calculated peak intensity from single-Gaussian fit; while the rest were set to $2R_{\text{input}}$. Statistically, after averaging over 200 generations, the calculated $R_{\text{obs}}$ had a systematic deviation from $R_{\text{input}}$. We demonstrate examples of the variations of calculated $R_{\text{obs}}$ as functions of SNR in Figure \ref{fig :anomaly_curve}(A). In this figure, the curves with different colors are averaged $R_{\text{obs}}$ for different $\delta\lambda_{\text{instr}}$. It is to be noted that these values are statistical averaged values over 200 generations in the Monte-Carlo simulation. From Figure \ref{fig :anomaly_curve} it is clear that the calculated $R_{\text{obs}}$ are decreasing monotonically with increasing SNR for each $\delta\lambda_{\text{instr}}$. However, due to the degradation of spectral resolution, systematic deviations were implemented on the calculated $R_{\text{obs}}$. This is reasonable because, with larger $\delta\lambda_{\text{instr}}$ the spectral line profile will be broadened, followed with a reduction in the peak intensity. In this case, as $\delta\lambda_{\text{instr}}$ increases, the obtained $R_{\text{obs}}$ will be smaller than $R_{\text{input}}$, leading to negative values of $R_{\text{obs}}-R_{\text{input}}$. The obtained $R_{\text{obs}}$ are averaged values over 200 generations, and we set the values to be $2R_{\text{input}}$ for those line profiles disobeying the 3-sigma criterion. For SNR and $\delta\lambda_{\text{instr}}$ values close to the critical curve, there will be more line profile generations that do not satsify the 3-sigma criterion, thus their $R_{\text{obs}}$ values are $2R_{\text{input}}$, we will then average more $2R_{\text{input}}$ than calculated $R_{\text{obs}}$. As a result, the statistical averaged $R_{\text{obs}}$ value for SNR and $\delta\lambda_{\text{instr}}$ near the critical curve will be larger than those with higher SNR and spectral resolutions. For this reason, although in Figure \ref{fig :anomaly_curve}(B), $(R_{\text{obs}}-R_{\text{input}})/R_{\text{input}}$ maintains the monotonically decreasing trend, its absolute value will experience a decrease followed by an increase due to the statistical deviation (Figure \ref{fig :anomaly_curve}(C)). Nevertheless, the absolute difference $\left|R_{\text{obs}}-R_{\text{input}}\right|/R_{\text{input}}$ approaches an asymptote as the SNR increases, with the minimum generally occurs near the critical curves. In other words, for a given $\delta\lambda_{\text{instr}}$, the calculated $R_{\text{obs}}$, as well as $v_{\text{obs}}$, are almost constant above the critical SNR, therefore, despite the statistical deviation on $R_{\text{obs}}$, we can still use the results above the critical curves to evaluate the accuracy of calculated CME peak intensity. 

\section{A Slow CME Example}\label{ap:ex}

\begin{figure}[htbp!]
\centering
\includegraphics[width=0.5\textwidth]{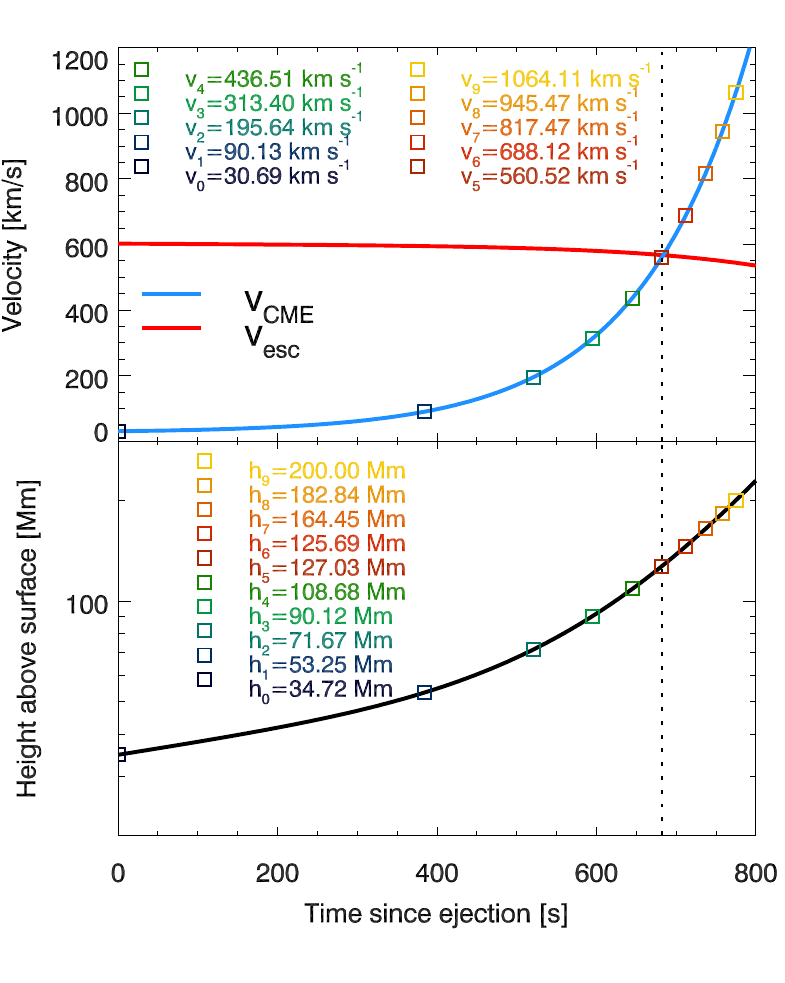}
\caption{Similar to Figure \ref{fig:f2_cme_esc_vel_h_new} but for a CME with slower acceleration.}  \label{fig:cme_esc_vel_h_H1}
\end{figure}

The velocity evolution profile used in previous analysis represents fast acceleration of a CME. For some CMEs, however, the acceleration is slower. We adopted different height and velocity profiles from \cite{2020ApJ...894...85C} (the sample H1). The two profiles describe the kinematic evolution of a slow solar CME, which are depicted in Figure \ref{fig:cme_esc_vel_h_H1}.

\begin{figure}[htbp!]
\centering
\includegraphics[width=0.95\textwidth]{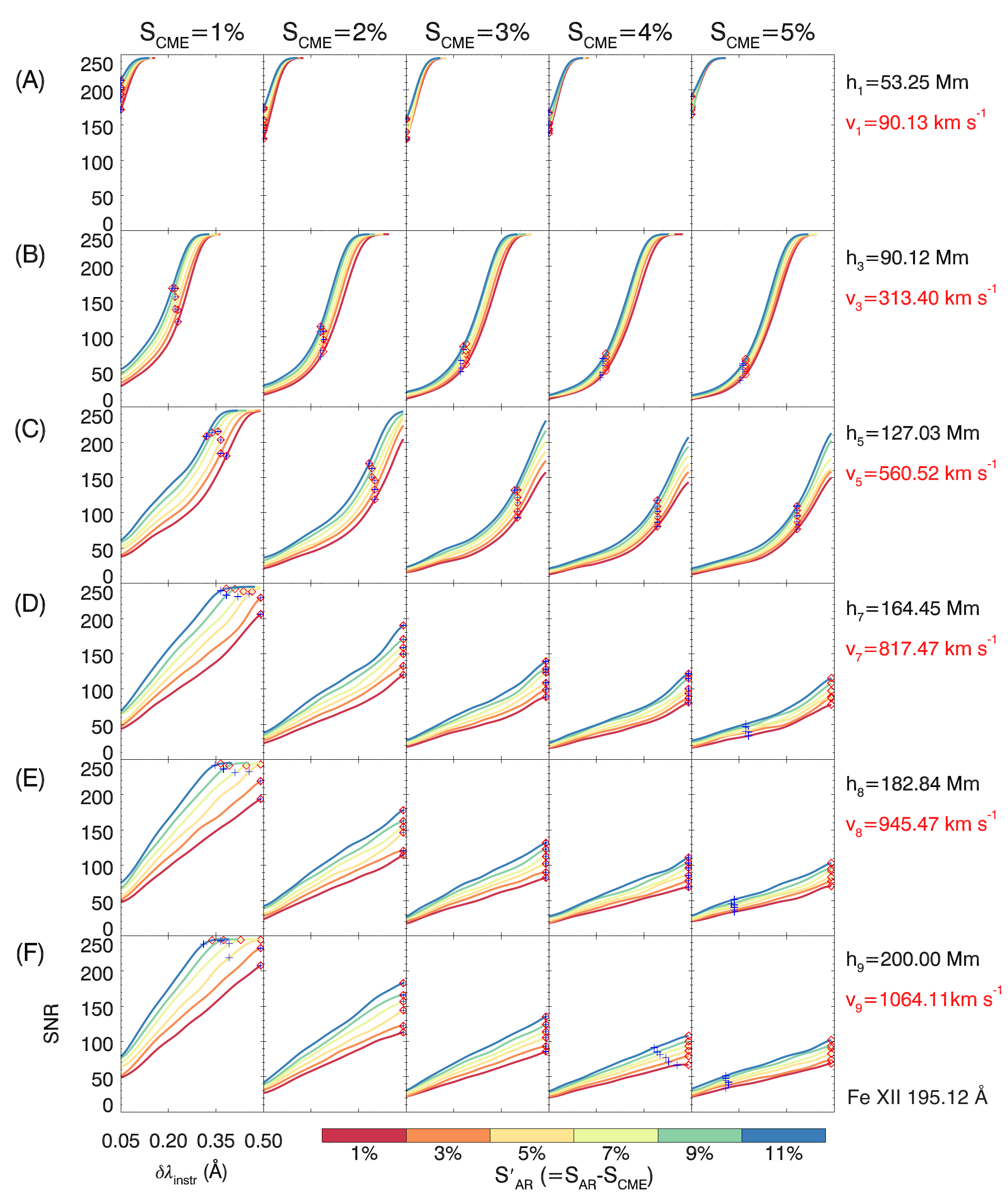}
\caption{Similar to Figure \ref{fig :detect_final_cme_curve} but for velocity and height profiles shown in Figure \ref{fig:cme_esc_vel_h_H1}. Note that we can hardly detect CME signals at the height of $h_1=53.25$ Mm with a speed of only around 90 $\text{km}\ \text{s}^{-1}$ (see row (A)) because the velocity is not large enough for the identification of blue wing enhancement. Here $S_{\text{CME}}$ and $S_{\text{AR}}$ are the initial areas at the height $h_0$.}  \label{fig :detect_final_cme_curve_H1}
\end{figure}

We applied the analytical expressions for full disk-integrated line profiles during CMEs and synthesized different profiles following our previous analyses. Similar evaluations on accuracy of calculated CME velocity and peak intensity from the synthetic profiles were also made, setting constraints on instrumental parameters. The overall constraints are summarized in Figure \ref{fig :detect_final_cme_curve_H1}. It is to be noted that, due to the slow acceleration, at lower heights (e.g., $h_0=34.72$ Mm or $h_1=53.25$ Mm) the velocity of the CME is too small for its spectral component to be identified in the spectra, thus we can hardly have parameter constraints at these heights. Under typical solar conditions, referring to the the second column of Figure \ref{fig :detect_final_cme_curve_H1}, if we want to detect a successfully erupted CME with its velocity exceeding the local escape velocity (e.g., $h_5=127.03$ Mm and $v_5=560.52\ \text{km}\ \text{s}^{-1}$), for Fe \sc{xii}\rm{} 195.12 \AA, the required minimum SNR is about 100-150. For the same $\delta\lambda_{\text{instr}}$, a higher SNR is required than in the case of the fast-accelerated CME. This is reasonable because, compared to the fast-accelerated case, the current CME travels to a higher height when it exceeds the escape velocity, where its emission is weaker. The minimum spectral resolving power is also $R\sim 500$ (see second column of row (C) in Figure \ref{fig :detect_final_cme_curve_H1}). As can be summarized from Figures \ref{fig :detect_final_cme_curve}, \ref{fig :detect_final_cme_curve_FeXV} and \ref{fig :detect_final_cme_curve_H1}, as long as the CME can be accelerated over the local escape velocity ($v_{\text{CME}}> 600\ \text{km}\ \text{s}^{-1}$), with a minimum spectral resolving power of $R_{\text{min}}\approx 500$ and a typical signal-to-noise ratio of $\sim 100$, the secondary component caused by CME eruptions can be well separated from the primary component contributed by background emissions. 

\bibliography{manuscript}{}

\begin{thebibliography}{}
\expandafter\ifx\csname natexlab\endcsname\relax\def\natexlab#1{#1}\fi
\providecommand{\url}[1]{\href{#1}{#1}}
\providecommand{\dodoi}[1]{doi:~\href{http://doi.org/#1}{\nolinkurl{#1}}}
\providecommand{\doeprint}[1]{\href{http://ascl.net/#1}{\nolinkurl{http://ascl.net/#1}}}
\providecommand{\doarXiv}[1]{\href{https://arxiv.org/abs/#1}{\nolinkurl{https://arxiv.org/abs/#1}}}

\bibitem[{{Argiroffi} {et~al.}(2019){Argiroffi}, {Reale}, {Drake},
  {Ciaravella}, {Testa}, {Bonito}, {Miceli}, {Orlando}, \&
  {Peres}}]{2019NatAs...3..742A}
{Argiroffi}, C., {Reale}, F., {Drake}, J.~J., {et~al.} 2019, Nature Astronomy,
  3, 742, \dodoi{10.1038/s41550-019-0781-4}

\bibitem[{{Aschwanden}(2009)}]{2009AnGeo..27.3275A}
{Aschwanden}, M.~J. 2009, Annales Geophysicae, 27, 3275,
  \dodoi{10.5194/angeo-27-3275-2009}

\bibitem[{{Aschwanden} \& {Gopalswamy}(2019)}]{2019ApJ...877..149A}
{Aschwanden}, M.~J., \& {Gopalswamy}, N. 2019, \apj, 877, 149,
  \dodoi{10.3847/1538-4357/ab1b39}

\bibitem[{{Balmaceda} {et~al.}(2020){Balmaceda}, {Vourlidas}, {Stenborg}, \&
  {St. Cyr}}]{2020SoPh..295..107B}
{Balmaceda}, L.~A., {Vourlidas}, A., {Stenborg}, G., \& {St. Cyr}, O.~C. 2020,
  \solphys, 295, 107, \dodoi{10.1007/s11207-020-01672-6}

\bibitem[{{Bein} {et~al.}(2013){Bein}, {Temmer}, {Vourlidas}, {Veronig}, \&
  {Utz}}]{2013ApJ...768...31B}
{Bein}, B.~M., {Temmer}, M., {Vourlidas}, A., {Veronig}, A.~M., \& {Utz}, D.
  2013, \apj, 768, 31, \dodoi{10.1088/0004-637X/768/1/31}

\bibitem[{{Brown} {et~al.}(2008){Brown}, {Feldman}, {Seely}, {Korendyke}, \&
  {Hara}}]{2008ApJS..176..511B}
{Brown}, C.~M., {Feldman}, U., {Seely}, J.~F., {Korendyke}, C.~M., \& {Hara},
  H. 2008, \apjs, 176, 511, \dodoi{10.1086/529378}

\bibitem[{{Brueckner} {et~al.}(1995){Brueckner}, {Howard}, {Koomen},
  {Korendyke}, {Michels}, {Moses}, {Socker}, {Dere}, {Lamy}, {Llebaria},
  {Bout}, {Schwenn}, {Simnett}, {Bedford}, \& {Eyles}}]{1995SoPh..162..357B}
{Brueckner}, G.~E., {Howard}, R.~A., {Koomen}, M.~J., {et~al.} 1995, \solphys,
  162, 357, \dodoi{10.1007/BF00733434}

\bibitem[{{Cheng} {et~al.}(2013){Cheng}, {Zhang}, {Ding}, {Olmedo}, {Sun},
  {Guo}, \& {Liu}}]{2013ApJ...769L..25C}
{Cheng}, X., {Zhang}, J., {Ding}, M.~D., {et~al.} 2013, \apjl, 769, L25,
  \dodoi{10.1088/2041-8205/769/2/L25}

\bibitem[{{Cheng} {et~al.}(2020){Cheng}, {Zhang}, {Kliem}, {T{\"o}r{\"o}k},
  {Xing}, {Zhou}, {Inhester}, \& {Ding}}]{2020ApJ...894...85C}
{Cheng}, X., {Zhang}, J., {Kliem}, B., {et~al.} 2020, \apj, 894, 85,
  \dodoi{10.3847/1538-4357/ab886a}

\bibitem[{{Cheng} {et~al.}(2011){Cheng}, {Zhang}, {Liu}, \&
  {Ding}}]{2011ApJ...732L..25C}
{Cheng}, X., {Zhang}, J., {Liu}, Y., \& {Ding}, M.~D. 2011, \apjl, 732, L25,
  \dodoi{10.1088/2041-8205/732/2/L25}

\bibitem[{{Cheng} {et~al.}(2012){Cheng}, {Zhang}, {Saar}, \&
  {Ding}}]{2012ApJ...761...62C}
{Cheng}, X., {Zhang}, J., {Saar}, S.~H., \& {Ding}, M.~D. 2012, \apj, 761, 62,
  \dodoi{10.1088/0004-637X/761/1/62}

\bibitem[{{Colaninno} \& {Vourlidas}(2006)}]{2006ApJ...652.1747C}
{Colaninno}, R.~C., \& {Vourlidas}, A. 2006, \apj, 652, 1747,
  \dodoi{10.1086/507943}

\bibitem[{{Corona-Romero} \& {Riley}(2020)}]{2020AnGeo..38..657C}
{Corona-Romero}, P., \& {Riley}, P. 2020, Annales Geophysicae, 38, 657,
  \dodoi{10.5194/angeo-38-657-2020}

\bibitem[{{Culhane} {et~al.}(2007){Culhane}, {Harra}, {James}, {Al-Janabi},
  {Bradley}, {Chaudry}, {Rees}, {Tandy}, {Thomas}, {Whillock}, {Winter},
  {Doschek}, {Korendyke}, {Brown}, {Myers}, {Mariska}, {Seely}, {Lang}, {Kent},
  {Shaughnessy}, {Young}, {Simnett}, {Castelli}, {Mahmoud}, {Mapson-Menard},
  {Probyn}, {Thomas}, {Davila}, {Dere}, {Windt}, {Shea}, {Hagood}, {Moye},
  {Hara}, {Watanabe}, {Matsuzaki}, {Kosugi}, {Hansteen}, \&
  {Wikstol}}]{2007SoPh..243...19C}
{Culhane}, J.~L., {Harra}, L.~K., {James}, A.~M., {et~al.} 2007, \solphys, 243,
  19, \dodoi{10.1007/s01007-007-0293-1}

\bibitem[{{De Pontieu} {et~al.}(2009){De Pontieu}, {McIntosh}, {Hansteen}, \&
  {Schrijver}}]{2009ApJ...701L...1D}
{De Pontieu}, B., {McIntosh}, S.~W., {Hansteen}, V.~H., \& {Schrijver}, C.~J.
  2009, \apjl, 701, L1, \dodoi{10.1088/0004-637X/701/1/L1}

\bibitem[{{de Wijn} {et~al.}(2012){de Wijn}, {Burkepile}, {Tomczyk}, {Nelson},
  {Huang}, \& {Gallagher}}]{2012SPIE.8444E..3ND}
{de Wijn}, A.~G., {Burkepile}, J.~T., {Tomczyk}, S., {et~al.} 2012, in Society
  of Photo-Optical Instrumentation Engineers (SPIE) Conference Series, Vol.
  8444, Ground-based and Airborne Telescopes IV, ed. L.~M. {Stepp},
  R.~{Gilmozzi}, \& H.~J. {Hall}, 84443N, \dodoi{10.1117/12.926511}

\bibitem[{{Del Zanna}(2012)}]{2012A&A...537A..38D}
{Del Zanna}, G. 2012, \aap, 537, A38, \dodoi{10.1051/0004-6361/201117592}

\bibitem[{{Del Zanna} {et~al.}(2021){Del Zanna}, {Dere}, {Young}, \&
  {Landi}}]{2021ApJ...909...38D}
{Del Zanna}, G., {Dere}, K.~P., {Young}, P.~R., \& {Landi}, E. 2021, \apj, 909,
  38, \dodoi{10.3847/1538-4357/abd8ce}

\bibitem[{{Del Zanna} \& {Mason}(2018)}]{2018LRSP...15....5D}
{Del Zanna}, G., \& {Mason}, H.~E. 2018, Living Reviews in Solar Physics, 15,
  5, \dodoi{10.1007/s41116-018-0015-3}

\bibitem[{{Dere} {et~al.}(1997){Dere}, {Landi}, {Mason}, {Monsignori Fossi}, \&
  {Young}}]{1997A&AS..125..149D}
{Dere}, K.~P., {Landi}, E., {Mason}, H.~E., {Monsignori Fossi}, B.~C., \&
  {Young}, P.~R. 1997, \aaps, 125, 149, \dodoi{10.1051/aas:1997368}

\bibitem[{{Dobrzycka} {et~al.}(2003){Dobrzycka}, {Raymond}, {Biesecker}, {Li},
  \& {Ciaravella}}]{2003ApJ...588..586D}
{Dobrzycka}, D., {Raymond}, J.~C., {Biesecker}, D.~A., {Li}, J., \&
  {Ciaravella}, A. 2003, \apj, 588, 586, \dodoi{10.1086/374047}

\bibitem[{Farmer(2005)}]{farmer2005volume}
Farmer, J. 2005, Australian Senior Mathematics Journal, 19, 49

\bibitem[{{Feng} {et~al.}(2015){Feng}, {Wang}, {Shen}, {Shen}, {Inhester},
  {Lu}, \& {Gan}}]{2015ApJ...812...70F}
{Feng}, L., {Wang}, Y., {Shen}, F., {et~al.} 2015, \apj, 812, 70,
  \dodoi{10.1088/0004-637X/812/1/70}

\bibitem[{{Fontenla} {et~al.}(1999){Fontenla}, {White}, {Fox}, {Avrett}, \&
  {Kurucz}}]{1999ApJ...518..480F}
{Fontenla}, J., {White}, O.~R., {Fox}, P.~A., {Avrett}, E.~H., \& {Kurucz},
  R.~L. 1999, \apj, 518, 480, \dodoi{10.1086/307258}

\bibitem[{{Gallagher} {et~al.}(2003){Gallagher}, {Lawrence}, \&
  {Dennis}}]{2003ApJ...588L..53G}
{Gallagher}, P.~T., {Lawrence}, G.~R., \& {Dennis}, B.~R. 2003, \apjl, 588,
  L53, \dodoi{10.1086/375504}

\bibitem[{{Giordano} {et~al.}(2013){Giordano}, {Ciaravella}, {Raymond}, {Ko},
  \& {Suleiman}}]{2013JGRA..118..967G}
{Giordano}, S., {Ciaravella}, A., {Raymond}, J.~C., {Ko}, Y.~K., \& {Suleiman},
  R. 2013, Journal of Geophysical Research (Space Physics), 118, 967,
  \dodoi{10.1002/jgra.50166}

\bibitem[{{Gopalswamy}(2016)}]{2016GSL.....3....8G}
{Gopalswamy}, N. 2016, Geoscience Letters, 3, 8,
  \dodoi{10.1186/s40562-016-0039-2}

\bibitem[{{Gopalswamy} {et~al.}(2014){Gopalswamy}, {Akiyama}, {Yashiro}, {Xie},
  {M{\"a}kel{\"a}}, \& {Michalek}}]{2014GeoRL..41.2673G}
{Gopalswamy}, N., {Akiyama}, S., {Yashiro}, S., {et~al.} 2014, \grl, 41, 2673,
  \dodoi{10.1002/2014GL059858}

\bibitem[{{Gopalswamy} {et~al.}(2009{\natexlab{a}}){Gopalswamy}, {Dal Lago},
  {Yashiro}, \& {Akiyama}}]{2009CEAB...33..115G}
{Gopalswamy}, N., {Dal Lago}, A., {Yashiro}, S., \& {Akiyama}, S.
  2009{\natexlab{a}}, Central European Astrophysical Bulletin, 33, 115

\bibitem[{{Gopalswamy} {et~al.}(2009{\natexlab{b}}){Gopalswamy}, {Yashiro},
  {Michalek}, {Stenborg}, {Vourlidas}, {Freeland}, \&
  {Howard}}]{2009EM&P..104..295G}
{Gopalswamy}, N., {Yashiro}, S., {Michalek}, G., {et~al.} 2009{\natexlab{b}},
  Earth Moon and Planets, 104, 295, \dodoi{10.1007/s11038-008-9282-7}

\bibitem[{{Houdebine} {et~al.}(1990){Houdebine}, {Foing}, \&
  {Rodono}}]{1990A&A...238..249H}
{Houdebine}, E.~R., {Foing}, B.~H., \& {Rodono}, M. 1990, \aap, 238, 249

\bibitem[{{Howard} \& {Vourlidas}(2018)}]{2018SoPh..293...55H}
{Howard}, R.~A., \& {Vourlidas}, A. 2018, \solphys, 293, 55,
  \dodoi{10.1007/s11207-018-1274-9}

\bibitem[{{Judge}(2020)}]{2020MNRAS.491..576J}
{Judge}, P.~G. 2020, \mnras, 491, 576, \dodoi{10.1093/mnras/stz3063}

\bibitem[{{Kaiser} {et~al.}(2008){Kaiser}, {Kucera}, {Davila}, {St. Cyr},
  {Guhathakurta}, \& {Christian}}]{2008SSRv..136....5K}
{Kaiser}, M.~L., {Kucera}, T.~A., {Davila}, J.~M., {et~al.} 2008, \ssr, 136, 5,
  \dodoi{10.1007/s11214-007-9277-0}

\bibitem[{{Landi}(2007)}]{2007A&A...476..675L}
{Landi}, E. 2007, \aap, 476, 675, \dodoi{10.1051/0004-6361:20077929}

\bibitem[{{Landi} {et~al.}(2016){Landi}, {Habbal}, \&
  {Tomczyk}}]{2016JGRA..121.8237L}
{Landi}, E., {Habbal}, S.~R., \& {Tomczyk}, S. 2016, Journal of Geophysical
  Research (Space Physics), 121, 8237, \dodoi{10.1002/2016JA022598}

\bibitem[{{Landi} {et~al.}(2010){Landi}, {Raymond}, {Miralles}, \&
  {Hara}}]{2010ApJ...711...75L}
{Landi}, E., {Raymond}, J.~C., {Miralles}, M.~P., \& {Hara}, H. 2010, \apj,
  711, 75, \dodoi{10.1088/0004-637X/711/1/75}

\bibitem[{{Lee} {et~al.}(2016){Lee}, {Moon}, {Lee}, {Lee}, \&
  {Kim}}]{2016JGRA..121.2853L}
{Lee}, J.-O., {Moon}, Y.~J., {Lee}, J.-Y., {Lee}, K.-S., \& {Kim}, R.~S. 2016,
  Journal of Geophysical Research (Space Physics), 121, 2853,
  \dodoi{10.1002/2015JA022321}

\bibitem[{{Leitzinger} {et~al.}(2011){Leitzinger}, {Odert}, {Ribas},
  {Hanslmeier}, {Lammer}, {Khodachenko}, {Zaqarashvili}, \&
  {Rucker}}]{2011A&A...536A..62L}
{Leitzinger}, M., {Odert}, P., {Ribas}, I., {et~al.} 2011, \aap, 536, A62,
  \dodoi{10.1051/0004-6361/201015985}

\bibitem[{{Lemen} {et~al.}(2012){Lemen}, {Title}, {Akin}, {Boerner}, {Chou},
  {Drake}, {Duncan}, {Edwards}, {Friedlaender}, {Heyman}, {Hurlburt}, {Katz},
  {Kushner}, {Levay}, {Lindgren}, {Mathur}, {McFeaters}, {Mitchell}, {Rehse},
  {Schrijver}, {Springer}, {Stern}, {Tarbell}, {Wuelser}, {Wolfson}, {Yanari},
  {Bookbinder}, {Cheimets}, {Caldwell}, {Deluca}, {Gates}, {Golub}, {Park},
  {Podgorski}, {Bush}, {Scherrer}, {Gummin}, {Smith}, {Auker}, {Jerram},
  {Pool}, {Soufli}, {Windt}, {Beardsley}, {Clapp}, {Lang}, \&
  {Waltham}}]{2012SoPh..275...17L}
{Lemen}, J.~R., {Title}, A.~M., {Akin}, D.~J., {et~al.} 2012, \solphys, 275,
  17, \dodoi{10.1007/s11207-011-9776-8}

\bibitem[{{Li} {et~al.}(2019){Li}, {Ding}, {Hong}, {Li}, \&
  {Gan}}]{2019ApJ...879...30L}
{Li}, Y., {Ding}, M.~D., {Hong}, J., {Li}, H., \& {Gan}, W.~Q. 2019, \apj, 879,
  30, \dodoi{10.3847/1538-4357/ab245a}

\bibitem[{{Low}(1984)}]{1984ApJ...281..392L}
{Low}, B.~C. 1984, \apj, 281, 392, \dodoi{10.1086/162110}

\bibitem[{{Mart{\'\i}nez-Sykora} {et~al.}(2011){Mart{\'\i}nez-Sykora}, {De
  Pontieu}, {Hansteen}, \& {McIntosh}}]{2011ApJ...732...84M}
{Mart{\'\i}nez-Sykora}, J., {De Pontieu}, B., {Hansteen}, V., \& {McIntosh},
  S.~W. 2011, \apj, 732, 84, \dodoi{10.1088/0004-637X/732/2/84}

\bibitem[{{Mason} {et~al.}(2014){Mason}, {Woods}, {Caspi}, {Thompson}, \&
  {Hock}}]{2014ApJ...789...61M}
{Mason}, J.~P., {Woods}, T.~N., {Caspi}, A., {Thompson}, B.~J., \& {Hock},
  R.~A. 2014, \apj, 789, 61, \dodoi{10.1088/0004-637X/789/1/61}

\bibitem[{{Michalek} {et~al.}(2009){Michalek}, {Gopalswamy}, \&
  {Yashiro}}]{2009SoPh..260..401M}
{Michalek}, G., {Gopalswamy}, N., \& {Yashiro}, S. 2009, \solphys, 260, 401,
  \dodoi{10.1007/s11207-009-9464-0}

\bibitem[{{Milligan} \& {Dennis}(2009)}]{2009ApJ...699..968M}
{Milligan}, R.~O., \& {Dennis}, B.~R. 2009, \apj, 699, 968,
  \dodoi{10.1088/0004-637X/699/2/968}

\bibitem[{{Milligan} {et~al.}(2006){Milligan}, {Gallagher}, {Mathioudakis},
  {Bloomfield}, {Keenan}, \& {Schwartz}}]{2006ApJ...638L.117M}
{Milligan}, R.~O., {Gallagher}, P.~T., {Mathioudakis}, M., {et~al.} 2006,
  \apjl, 638, L117, \dodoi{10.1086/500555}

\bibitem[{{Namekata} {et~al.}(2021){Namekata}, {Maehara}, {Honda}, {Notsu},
  {Okamoto}, {Takahashi}, {Takayama}, {Ohshima}, {Saito}, {Katoh}, {Tozuka},
  {Murata}, {Ogawa}, {Niwano}, {Adachi}, {Oeda}, {Shiraishi}, {Isogai}, {Seki},
  {Ishii}, {Ichimoto}, {Nogami}, \& {Shibata}}]{2021NatAs...6..241N}
{Namekata}, K., {Maehara}, H., {Honda}, S., {et~al.} 2021, Nature Astronomy, 6,
  241, \dodoi{10.1038/s41550-021-01532-8}

\bibitem[{{Narechania} {et~al.}(2021){Narechania}, {Nikoli{\'c}}, {Freret}, {De
  Sterck}, \& {Groth}}]{2021JSWSC..11....8N}
{Narechania}, N.~M., {Nikoli{\'c}}, L., {Freret}, L., {De Sterck}, H., \&
  {Groth}, C. P.~T. 2021, Journal of Space Weather and Space Climate, 11, 8,
  \dodoi{10.1051/swsc/2020068}

\bibitem[{{Owens} {et~al.}(2017){Owens}, {Lockwood}, \&
  {Barnard}}]{2017NatSR...7.4152O}
{Owens}, M.~J., {Lockwood}, M., \& {Barnard}, L.~A. 2017, Scientific Reports,
  7, 4152, \dodoi{10.1038/s41598-017-04546-3}

\bibitem[{{Pant} {et~al.}(2021){Pant}, {Majumdar}, {Patel}, {Chauhan},
  {Banerjee}, \& {Gopalswamy}}]{2021FrASS...8...73P}
{Pant}, V., {Majumdar}, S., {Patel}, R., {et~al.} 2021, Frontiers in Astronomy
  and Space Sciences, 8, 73, \dodoi{10.3389/fspas.2021.634358}

\bibitem[{{Peter}(2010)}]{2010A&A...521A..51P}
{Peter}, H. 2010, \aap, 521, A51, \dodoi{10.1051/0004-6361/201014433}

\bibitem[{{Phillips} {et~al.}(2012){Phillips}, {Feldman}, \&
  {Landi}}]{2012uxss.book.....P}
{Phillips}, K. J.~H., {Feldman}, U., \& {Landi}, E. 2012, {Ultraviolet and
  X-ray Spectroscopy of the Solar Atmosphere}

\bibitem[{{Rivera} {et~al.}(2019){Rivera}, {Landi}, {Lepri}, \&
  {Gilbert}}]{2019ApJ...874..164R}
{Rivera}, Y.~J., {Landi}, E., {Lepri}, S.~T., \& {Gilbert}, J.~A. 2019, \apj,
  874, 164, \dodoi{10.3847/1538-4357/ab0e11}

\bibitem[{{Shen} {et~al.}(2012){Shen}, {Wu}, {Feng}, \&
  {Wu}}]{2012JGRA..11711101S}
{Shen}, F., {Wu}, S.~T., {Feng}, X., \& {Wu}, C.-C. 2012, Journal of
  Geophysical Research (Space Physics), 117, A11101,
  \dodoi{10.1029/2012JA017776}

\bibitem[{{St. Cyr} {et~al.}(2017){St. Cyr}, {Posner}, \&
  {Burkepile}}]{2017SpWea..15..240S}
{St. Cyr}, O.~C., {Posner}, A., \& {Burkepile}, J.~T. 2017, Space Weather, 15,
  240, \dodoi{10.1002/2016SW001545}

\bibitem[{{Thernisien} {et~al.}(2006){Thernisien}, {Howard}, \&
  {Vourlidas}}]{2006ApJ...652..763T}
{Thernisien}, A.~F.~R., {Howard}, R.~A., \& {Vourlidas}, A. 2006, \apj, 652,
  763, \dodoi{10.1086/508254}

\bibitem[{{Tian} {et~al.}(2011){Tian}, {McIntosh}, {De Pontieu},
  {Mart{\'\i}nez-Sykora}, {Sechler}, \& {Wang}}]{2011ApJ...738...18T}
{Tian}, H., {McIntosh}, S.~W., {De Pontieu}, B., {et~al.} 2011, \apj, 738, 18,
  \dodoi{10.1088/0004-637X/738/1/18}

\bibitem[{{Tian} {et~al.}(2012){Tian}, {McIntosh}, {Xia}, {He}, \&
  {Wang}}]{2012ApJ...748..106T}
{Tian}, H., {McIntosh}, S.~W., {Xia}, L., {He}, J., \& {Wang}, X. 2012, \apj,
  748, 106, \dodoi{10.1088/0004-637X/748/2/106}

\bibitem[{{Tian} {et~al.}(2013){Tian}, {Tomczyk}, {McIntosh}, {Bethge}, {de
  Toma}, \& {Gibson}}]{2013SoPh..288..637T}
{Tian}, H., {Tomczyk}, S., {McIntosh}, S.~W., {et~al.} 2013, \solphys, 288,
  637, \dodoi{10.1007/s11207-013-0317-5}

\bibitem[{{Tian} {et~al.}(2015){Tian}, {Young}, {Reeves}, {Chen}, {Liu}, \&
  {McKillop}}]{2015ApJ...811..139T}
{Tian}, H., {Young}, P.~R., {Reeves}, K.~K., {et~al.} 2015, \apj, 811, 139,
  \dodoi{10.1088/0004-637X/811/2/139}

\bibitem[{{Tomczyk} {et~al.}(2008){Tomczyk}, {Card}, {Darnell}, {Elmore},
  {Lull}, {Nelson}, {Streander}, {Burkepile}, {Casini}, \&
  {Judge}}]{2008SoPh..247..411T}
{Tomczyk}, S., {Card}, G.~L., {Darnell}, T., {et~al.} 2008, \solphys, 247, 411,
  \dodoi{10.1007/s11207-007-9103-6}

\bibitem[{{Vida} {et~al.}(2019){Vida}, {Leitzinger}, {Kriskovics}, {Seli},
  {Odert}, {Kov{\'a}cs}, {Korhonen}, \& {van
  Driel-Gesztelyi}}]{2019A&A...623A..49V}
{Vida}, K., {Leitzinger}, M., {Kriskovics}, L., {et~al.} 2019, \aap, 623, A49,
  \dodoi{10.1051/0004-6361/201834264}

\bibitem[{{Vourlidas} \& {Howard}(2006)}]{2006ApJ...642.1216V}
{Vourlidas}, A., \& {Howard}, R.~A. 2006, \apj, 642, 1216,
  \dodoi{10.1086/501122}

\bibitem[{{Wang} {et~al.}(2009){Wang}, {Zhang}, \&
  {Shen}}]{2009JGRA..11410104W}
{Wang}, Y., {Zhang}, J., \& {Shen}, C. 2009, Journal of Geophysical Research
  (Space Physics), 114, A10104, \dodoi{10.1029/2009JA014360}

\bibitem[{{Woods} {et~al.}(2012){Woods}, {Eparvier}, {Hock}, {Jones},
  {Woodraska}, {Judge}, {Didkovsky}, {Lean}, {Mariska}, {Warren}, {McMullin},
  {Chamberlin}, {Berthiaume}, {Bailey}, {Fuller-Rowell}, {Sojka}, {Tobiska}, \&
  {Viereck}}]{2012SoPh..275..115W}
{Woods}, T.~N., {Eparvier}, F.~G., {Hock}, R., {et~al.} 2012, \solphys, 275,
  115, \dodoi{10.1007/s11207-009-9487-6}

\bibitem[{{Xu} {et~al.}(2022){Xu}, {Tian}, {Hou}, \& {et al.}}]{Xu2022}
{Xu}, Y., {Tian}, H., {Hou}, Z., \& {et al.} 2022, submitted to ApJ

\bibitem[{{Young} {et~al.}(2007){Young}, {Del Zanna}, {Mason}, {Dere}, {Landi},
  {Landini}, {Doschek}, {Brown}, {Culhane}, {Harra}, {Watanabe}, \&
  {Hara}}]{2007PASJ...59S.857Y}
{Young}, P.~R., {Del Zanna}, G., {Mason}, H.~E., {et~al.} 2007, \pasj, 59,
  S857, \dodoi{10.1093/pasj/59.sp3.S857}

\bibitem[{{Zhang} {et~al.}(2001){Zhang}, {Dere}, {Howard}, {Kundu}, \&
  {White}}]{2001ApJ...559..452Z}
{Zhang}, J., {Dere}, K.~P., {Howard}, R.~A., {Kundu}, M.~R., \& {White}, S.~M.
  2001, \apj, 559, 452, \dodoi{10.1086/322405}

\bibitem[{{Zhang} {et~al.}(2004){Zhang}, {Dere}, {Howard}, \&
  {Vourlidas}}]{2004ApJ...604..420Z}
{Zhang}, J., {Dere}, K.~P., {Howard}, R.~A., \& {Vourlidas}, A. 2004, \apj,
  604, 420, \dodoi{10.1086/381725}

\bibitem[{{Zhou} {et~al.}(2003){Zhou}, {Wang}, \& {Cao}}]{2003A&A...397.1057Z}
{Zhou}, G., {Wang}, J., \& {Cao}, Z. 2003, \aap, 397, 1057,
  \dodoi{10.1051/0004-6361:20021463}

\end{thebibliography}
\bibliographystyle{aasjournal}


\end{CJK*}
\end{document}